\documentclass{aastex}
\usepackage{spr-astr-addons}
\usepackage{url}
\usepackage{longtable}
\usepackage{tabularx}

    \usepackage{epsfig}

\RequirePackage{color}

\newcommand{\emaila}{}

\begin{document}

\title{Virilization of the Broad Line Region in Active Galactic Nuclei -- connection between shifts and widths of broad emission lines}
%%\slugcomment{Not to appear in Nonlearned J., 45.}
%% Running heads
\shorttitle{Virilization of the BLR in AGN -- connection between shifts and widths of broad emission lines}
\shortauthors{Jonic et al.}

\author{S. Joni\'c\altaffilmark{1}} \and \author{J. Kova\v cevi\' c-Doj\v cinovi\' c \altaffilmark{1}}
\affil{Astronomical Observatory, Volgina 7, 11060 Belgrade, Serbia}

\and
\author{D. Ili\' c\altaffilmark{2}} \and \author{L. \v C. Popovi\'c\ \altaffilmark{1,2}}
\affil{Department of Astronomy, Faculty of Mathematics, University of Belgrade, Studentski trg 16, 11000 Belgrade, Serbia}

\emaila{L.\v C. Popovi\'c: lpopovic@aob.bg.ac.rs}
\\
%\altaffiltext{1}{First Alternate Affilation.}
%\altaffiltext{2}{Second Alternate Affilation.}
%\altaffiltext{3}{Third Alternate Affilation.}

\begin{abstract}

We investigate the virilization of the emission lines ${\mathrm H\beta}$ and \ion{Mg}{2} in the sample of $287$ Type 1 Active Galactic Nuclei taken from the Sloan Digital Sky Survey database. 
We explore the connections between the intrinsic line shifts and full widths at different levels of maximal intensity. We found that:
(i) ${\mathrm H\beta}$ seems to be a good virial estimator of black hole masses, and an intrinsic redshift of ${\mathrm H\beta}$ is dominantly caused by the gravitational effect,
(ii) there is an anti-correlation between the redshift and width of the wings of the \ion{Mg}{2} line, (iii) the broad \ion{Mg}{2} line can be used as virial estimator only at 50\%  of the maximal intensity, while the widths and intrinsic shifts
of the line wings can not be used for this purpose.

\end{abstract}


%\section*{}
%\label{sec:intro}

\section{Introduction}

\indent 

\indent It is widely accepted that the center of Active Galactic Nuclei (AGN) consists of a supermassive black hole (BH) surrounded by an accretion disk gravitationally bound to the BH.
The accretion disk  is surrounded by an optically thick gaseous region - the Broad Line Region (BLR) that emits the Broad Emission Lines (BELs) \citep{Osterbrock1989,Krolik1999,Peterson2003,Popovic2006,Popovic2007,Ilic2012}, observed in Type 1 AGN.

\indent The BLR is ionized by the continuum radiation from the accretion disk and is influenced by the gravitational field of the central BH.  This region most likely has a complex structure \citep{Brotherton1994, Corbin1996, Sulentic2000, Popovic2004, Ilic2006, Hu2008}. 
 According to the one possible model of the BLR, the so-called two-component model, this emission region probably consists of two kinematically and geometrically different regions - the Very Broad Line Region (VBLR) that contributes to the broad line wings and which is closer to the BH, and the Intermediate Line Region (ILR), in which the broad line core is formed and which is further away from the BH \citep[etc]{Sulentic2000, Popovic2004, Bon2006, Hu2008, Bon2009a, Bon2009b, Kovacevic2010}. 
One theoretical explanation for this decomposition can be the contribution of the disk emission to the line wings \citep[see e.g.][]{Popovic2004, Bon2009a}, where the VBLR component represent the rough approximation of the disk emission.  However, there are also other possible models of the BLR with different geometry \citep[see e.g.][]{Netzer2010}, and here we choose to use the two-component model.

\indent There are several methods that can be used to estimate central BH mass in 
a galaxy \citep[for review see][]{Marziani2012, Shen2013, Peterson2014, Ilic2014}.
 For Type 1 AGN, the most appropriate methods for the BH mass estimation are those which use the strong broad emission lines, as e.g. the reverberation mapping \citep[][etc]{Blandford1982, Koratkar1991, Kaspi2000, Kaspi2005, Peterson2004}.

  This method is based on the monitoring of the variations of the continuum and BEL fluxes. The time lag between these variations can be measured and in that way, the  photometric radius of the BLR can be estimated \citep[see e.g.][]{Shapovalova2009,Peterson2014}. Then, the mass of the BH can be calculated using this radius and the width of the line, which gives the velocity of the BLR, assuming that the BLR gas is virially bounded to the 
   BH \citep{Gaskell1988}. Similarly, $R - L$ relationship \citep[]{Wandel1999, Kaspi2000,Bentz2006}, which is one of the outcomes of the reverberation mapping, enables the estimation of the photometric radius, and thus of the BH mass, from only one epoch spectrum \citep[see][]{Laor1998, Vestergaard2006, Kollmeier2006}.

The basic assumption of these methods is that the BLR gas is virilized, i.e. that the main broadening mechanism is the Keplerian motion around the supermassive BH. Taking into account that the BLR geometry can be complex \citep[see e.g.][etc]{Sulentic2000, Popovic2004, Gaskell2009}, there are many open questions that can be very important for the applicability of those methods. The most important questions are whether the virial assumption is correct for all BELs used in these methods, and is this assumption correct in general for all AGN population?

There is an evidence supporting the virial assumption in at least several AGN for which reverberation mapping have been performed \citep[][etc]{Gaskell1988, Kollatschny2003a, Peterson2004, Bentz2010, Afanasiev2015}. 
Some results imply that the virial relationship is tighter for  velocity dispersion of emission lines, than for the Full Width at Half Maximum (FWHM), since the FWHM is more influenced by the line asymmetry \citep{Peterson2004,Collin2006}. 

The lines which are the most frequently used as virial estimators are ${\mathrm H\beta}$ and \ion{Mg}{2} $\lambda\ 2800\ {\mathrm \AA}$  \citep[see][]{Marziani2012}. The ${\mathrm H\beta}$ line is present in optical part of AGN spectra and it has smaller asymmetries and shifts than other BELs, which is the reason why ${\mathrm H\beta}$ is mostly discussed as a BH mass estimator. At larger redshifts, $z \approx 3.7$, the ${\mathrm H\beta}$ line is redshifted to the IR part of the spectrum. Therefore, in order to estimate BH masses, one needs prominent BELs in the UV part,  
 that are, at higher redshifts, moved to the optical. The two strongest and mostly used UV lines are \ion{Mg}{2} $\lambda\ 2800\ {\mathrm \AA}$ and \ion{C}{4} $\lambda\ 1549\ {\mathrm \AA}$ \citep[]{McLure2002}. It seems that the \ion{C}{4} line is actually not a safe virial estimator of BH masses \citep[]{Netzer2007,Sulentic2007,Marziani2012}, because \ion{C}{4} often shows blueshifts and asymmetries that are interpreted as winds or outflows \citep[]{Gaskell1982, Gaskell2013, Marziani2013a}. \cite{Marziani2013a} showed that the \ion{Mg}{2} line can be used for BH mass estimations in case of 80-90\% high redshift quasars. The reverberation mapping is done only for ${\mathrm H\beta}$, but the \ion{Mg}{2} width can be used as a substitute for the ${\mathrm H\beta}$ width, with certain calibrations \citep[see][]{Vestergaard2009, Shen2012, Trakhtenbrot2012}.
However, both lines, ${\mathrm H\beta}$ and \ion{Mg}{2} show complex profiles, which must be taken into account, if they are used as virial estimators \citep[see][]{Marziani2012, Marziani2013a}. 
\citet{Marziani2012} pointed out that ${\mathrm H\beta}$ in a fraction of AGN with a large FWHM ${\mathrm H\beta}$ (Population B, FWHM ${\mathrm H\beta}$ $>$ 4000 km/s) is not the good virial estimator because of the redshifted VBLR component, while
\citet{Marziani2013b} found that a fraction of AGN with small FWHM ${\mathrm H\beta}$ (Population A), may have a blueshifted, non-virial component in \ion{Mg}{2}, probably caused by an outflow.

On the other hand, one more method, rarely used for the BH mass estimation, is based on the gravitational redshift of the BELs, which should be an indicator of the BH gravity \citep[see][]{Zheng1990, Popovic1995, Kollatschny2003b}. 
Since the BLR is stratified, the broad line wings are probably arising
in the deeper parts of the BLR, closer to the BH. It is possible that, at least in the line wings, the relativistic effect could be strong enough to be measured.
\citet{Zheng1990} proposed that the systemic redshift of the ${\mathrm H\beta}$ wings, relative to the line core, seen in the Population B AGN, can be caused by the gravitational redshift.

Here we investigate the virilization in the BLR, exploring the ${\mathrm H\beta}$ and \ion{Mg}{2} line widths at 5\%, 10\% and 50\% of the intensity maximum and corresponding shifts of their centroids with respect to the broad line peak. We are starting from
the hypothesis that if there is a dominant Keplerian motion affecting the line widths, there should exists the gravitational redshift in the line profile. We use the statistics of the large AGN 
sample to check the relationships between the widths and the shifts within the lines.

\bigskip

The paper is organized as following: In Section 2 we present our assumptions about the gas virilization, in Section 3 we describe the sample selection, spectra
decomposition and analysis. The results are given in Section 4, and discussed in Section 5. Finally, in Section 6 we outline our conclusions.

\section{The virilization in the emission gas}

\indent If the emitting gas is virilized, it is expected that the dominant line broadening is due to the Keplerian motion of the gas. 
Also, if the mass of the central source is large enough, the gravitational redshift is expected to be seen in BELs, or at least in the line wings, since the line wings are originating closer to the BH,
so the influence of the BH gravity is stronger. If the emission gas in the BLR is virilized, one can expect to observe correlations between the widths and gravitational redshifts of the BELs.

\indent From the virial theorem, the mass of the BH can be estimated as {\citep{Peterson2004}:

$$M = f\ \frac{R\ {\upsilon}^2}{G}, \eqno(1)$$

\noindent where $G$ is the gravitational constant, $R$ is the BLR radius, $\upsilon$ is the velocity of the gas in the BLR, and $f$ is the virial factor which depends on the BLR geometry and inclination.
The FWHM of the BELs can be used to measure the  mean gas velocity ($\upsilon$) in the BLR.

On the other hand, the redshift caused by the gravitational field, can give us an upper limit of the BH mass as \citep{Zheng1990}:

$$M = \frac{c^2 R}{G}z_G, \eqno(2)$$

\noindent where $ z_G$ is the gravitational redshift of the emission line.

\indent Assuming that in both equations (1) and (2)  $R$  represent the photometric radius, we can obtain the expected relationship between $z_G$ and FWHM of the line:

$$z_G \sim FWHM^2. \eqno(3)$$

\indent Therefore, if the virial assumption is valid, we can expect that there is a  power law dependence between the gravitational redshift and gas velocity, with exponent n = 2. For FWHM, the corresponding $z_G$ is ${\Delta z}_{50}$, that is intrinsic redshift of the line, measured as a centroid shift at half maximum with respect to the broad line peak (for more details see Sec 3.3). Consequently, there should be a linear correlation between logarithms of FWHM and ${\Delta z}_{50}$, $log ({\Delta z}_{50}) = C\ log (FWHM)$, where $C$ is a constant. Both parameters are given in km/s.

 Note that at higher velocities (widths at  10 \% and 5 \% of line maximum), we also expect a relationship like equation (1) although this has not been proven observationally. We do not have to know $f$ for these velocities, only to assume it is the same for all objects where the BLR is virialized. Therefore, we expect the same relationship (equation 3) of these widths with corresponding intrinsic redshifts measured at 10 \% and 5 \% of line maximum.

\section{Sample and Analysis}

\subsection{The AGN Sample}

\indent For this research we used spectra from the Sloan Digital Sky Survey (SDSS) database, Data Release 7. We choose the sample with the broad ${\mathrm H\beta}$ and \ion{Mg}{2} $\lambda2800\ {\mathrm \AA}$ lines (their equivalent widths requested to be larger than zero) in the redshift range $0.407 \leq z \leq 0.643$. 
This sample of 293 AGN is initially used for research of the relations between the UV and optical \ion{Fe}{2} emission in the Type 1 AGN, so the detailed search criteria and fitting procedures are given in \citet{Kovacevic2015}. After eliminating the spectra with very noisy UV part, near the \ion{Mg}{2} line, our sample contained 287 AGN spectra.

Since in this investigation we search for the gravitational redshift in the lines, we measure the intrinsic shifts in the whole sample (see Section 3.3), and we eliminate all spectra with blueshifted ${\mathrm H\beta}$/\ion{Mg}{2} profiles. In these spectra, some other effects (as e.g.\ outflows), could be more dominant than the gravitational effects, so they are not convenient for this research. 
Therefore, we further investigate one subsample of 209 spectra in which ${\mathrm H\beta}$ lines have no blueshift, and another subsample of 150 spectra in which there is no blueshift in the \ion{Mg}{2} lines. Finally, to compare the properties between these two lines, we use a third subsample of 123 spectra in which both lines have no blueshift. We should note here that with this selection, the large number of spectra are rejected from the initial sample. However, it is difficult to state that we have completely eliminated all spectra with strong outflow influence, since it is possible that the combination of the outflow and gravitational redshift produce symmetrical line shape.

\subsection{Line Fitting Procedure}

\indent In order to test the virialization in the broad ${\mathrm H\beta}$ and \ion{Mg}{2} $\lambda\ 2800\ {\mathrm \AA}$ lines, it is very important to remove all overlapping lines, and to obtain the clear BELs profiles.

 The complex shapes of BELs profiles indicate that different parts of BELs are coming from different emission regions. Since we assume that the Doppler broadening, caused by the motion of the emitting clouds, dominates in the line profiles of AGNs, we model the emission lines with the sum of Gaussians, where we suppose that each Gaussian represents the emission from kinematically different emission region. Therefore, the kinematical and physical properties of each emission region are reflected in the shifts, widths and intensities of Gaussians.
In the case of the broad Balmer lines, we find that the best fit is obtained using the two Gaussians \citep[see e.g.][]{Bon2009a, Kovacevic2010}, which supports the two-component model of the BLR \citep{Popovic2004, Bon2006, Hu2008}. The detailed description of this multi-Gaussian model and the fitting procedure is given in \citet{Kovacevic2015}, where fitting of this sample is performed in two spectral ranges: UV (2650-3050) and 
optical (4000-5500).

In the case of the optical range which includes the ${\mathrm H\beta}$ line, we subtract the continuum emission using the continuum windows given in \citet{Kuraszkiewicz2002}. 
Since the ${\mathrm H\beta}$ line overlaps with the optical \ion{Fe}{2} lines, \ion{He}{2} and [\ion{O}{3}]$\lambda\lambda$ 4959, 5007 \AA \ narrow lines, we apply the multi-Gaussian fitting model \citep[see Fig 2. in][]{Kovacevic2010}, 
to extract only the broad profile of ${\mathrm H\beta}$. 
The ${\mathrm H\beta}$ line is fi\-tted with three Gaussians: one represents the emission from the NLR,
and other two from the BLR. 
We assume that the Gaussian which fits the line wings represents the emission from the parts of the BLR closer to the BH (VBLR) and the one which fits the line core, 
the emission from the outer parts of the BLR (ILR). 
The [\ion{O}{3}] lines have the same width and shift as the ${\mathrm H\beta}$ narrow component, because we assume they all originate in the NLR. Additionally, the [\ion{O}{3}] lines have one more component which describes the asymmetry in the wings of these lines.
Numerous optical \ion{Fe}{2} lines are fitted with the \ion{Fe}{2} template\footnote{http://servo.aob.rs/FeII\_AGN/} presented in \citet{Kovacevic2010} and \citet{Shapovalova2012}.

In the case of \ion{Mg}{2}, the most challenging is to remove the numerous UV \ion{Fe}{2} lines which overlap with \ion{Mg}{2}. In order to fit the UV part of the spectra,
first the UV Balmer pseudo-continuum has to be subtracted. We use the UV Balmer pseudo-continuum model presented in \citet{Kovacevic2014}.
After that, we fit simultaneously the UV \ion{Fe}{2} lines in the range 2650-3050 \AA \ with the \ion{Mg}{2} line. For the UV \ion{Fe}{2} lines we use the model described in \citet{Popovic2003} and \citet{Kovacevic2015}. 
Note that the \ion{Mg}{2} line is 
the resonant doublet \ion{Mg}{2}$\lambda\lambda$  2795, 2803 \AA.  However, it can not be resolved since the components are very broad and overlap. Therefore, we fit the \ion{Mg}{2} line as a single line, with two Gaussian components, one which fits the core, and one which fits the wings of the line. 
The example of the spectrum decomposition near ${\mathrm H\beta}$ and \ion{Mg}{2} are given in Figure \ref{fig:Figure1}.

 There are various BLR models which predict the use of different functions, as e.g. Lorentzians or power-law functions, for fitting the emission line profiles in AGNs. In this work, we use the multi-Gaussian fitting decomposition only to remove
the overlapping lines and to reproduce the broad line profile of ${\mathrm H\beta}$ and \ion{Mg}{2}, which are obtained as the sum of the core and wing broad Gaussian.
 Since the parameters of this fitting decomposition are not used furthermore in this investigation, the results are not strongly affected with the function used for the line profile fitting.

\begin{figure}[h!]
    \centering
    \includegraphics[width=0.45\textwidth]{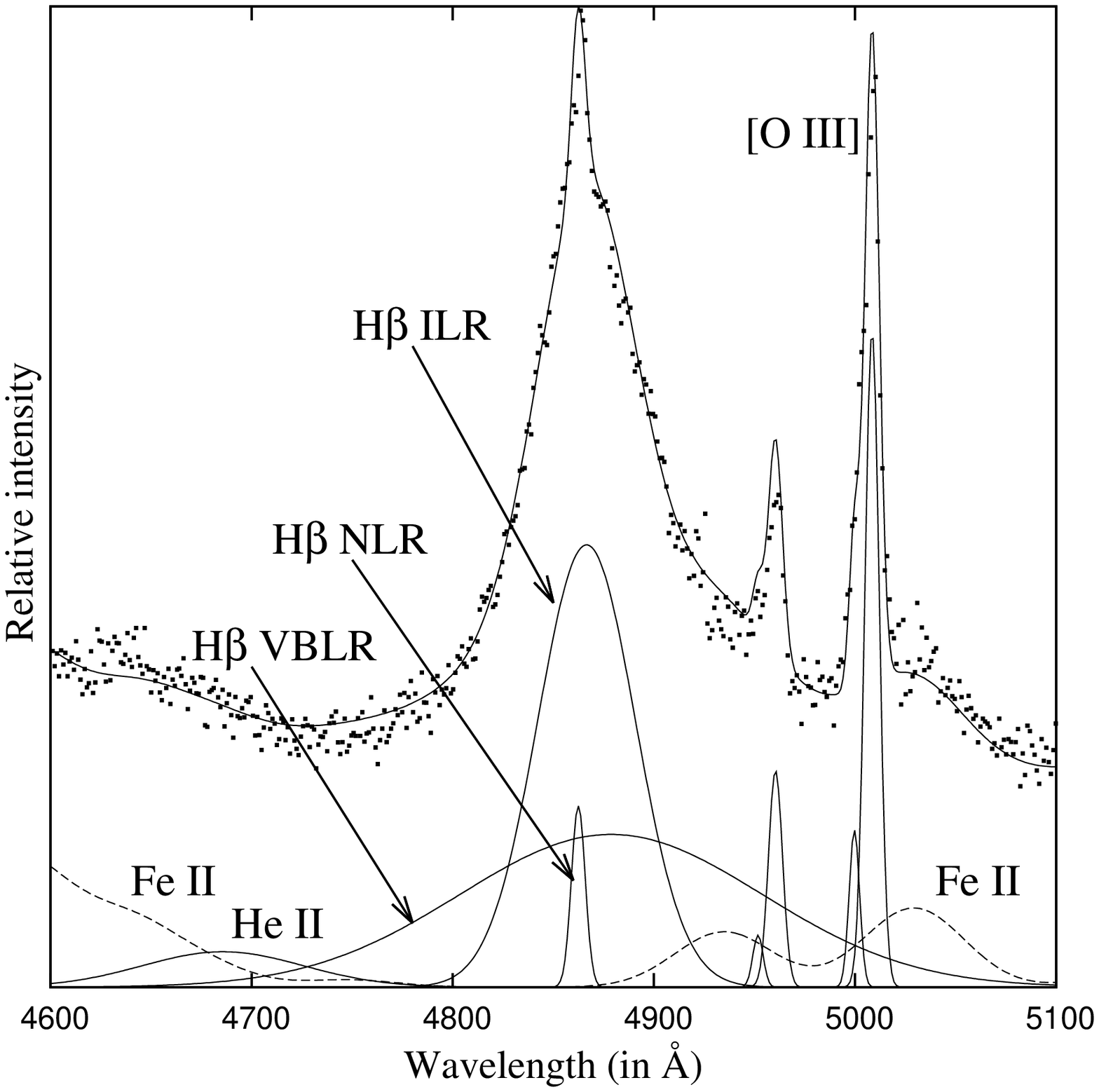}
 \includegraphics[width=0.45\textwidth]{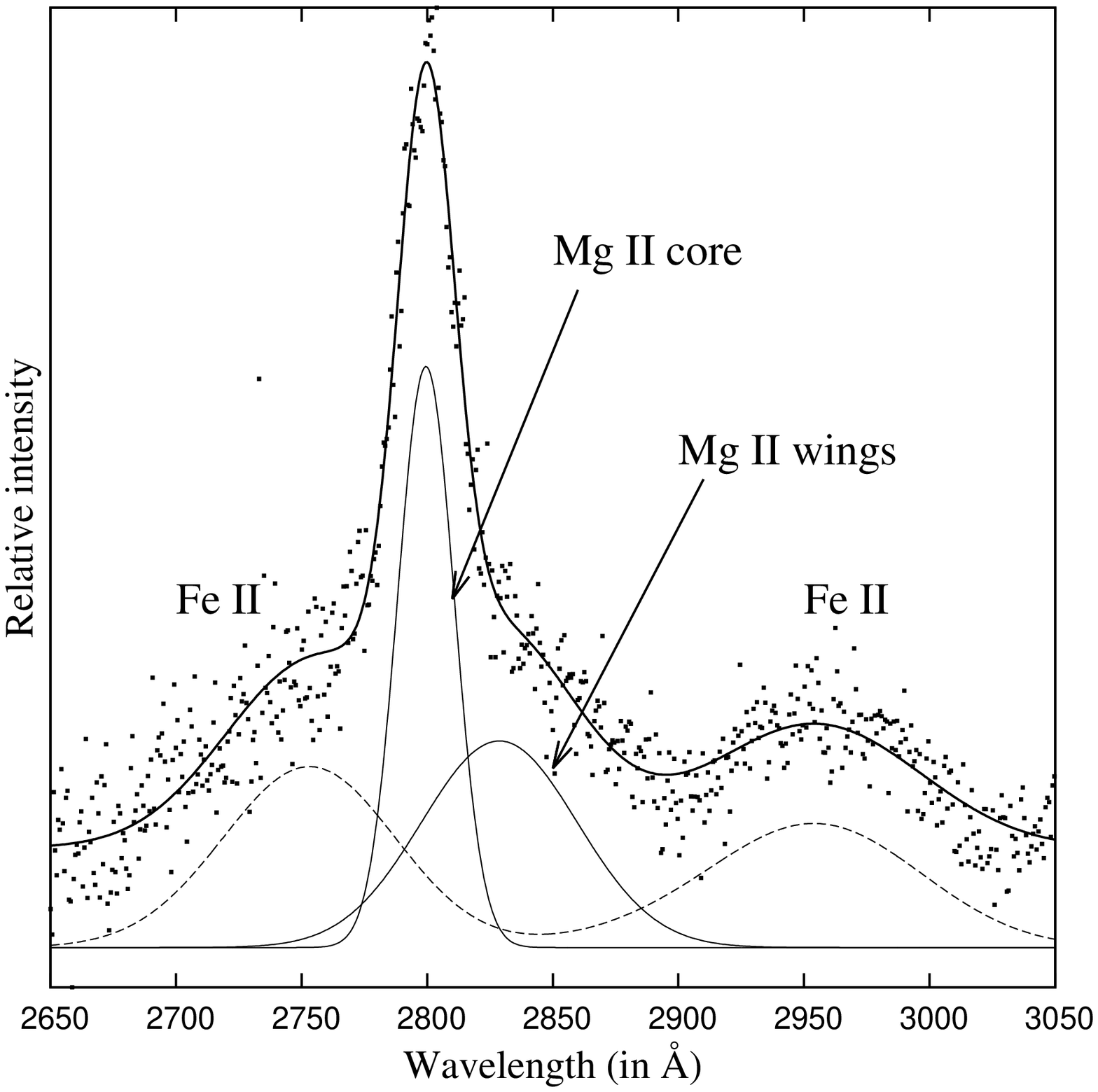}
    \caption{The example of the spectra decomposition for object SDSS J032205.04$+$001201.4, near ${\mathrm H\beta}$ (top) and \ion{Mg}{2} line (bottom). The broad ${\mathrm H\beta}$ and \ion{Mg}{2} lines are decomposed into two components which fit the line core and the line wings. The iron lines are denoted with dashed line.}
    \label{fig:Figure1}
\end{figure}

\begin{figure}[h!]
    \centering
    \includegraphics[width=0.45\textwidth]{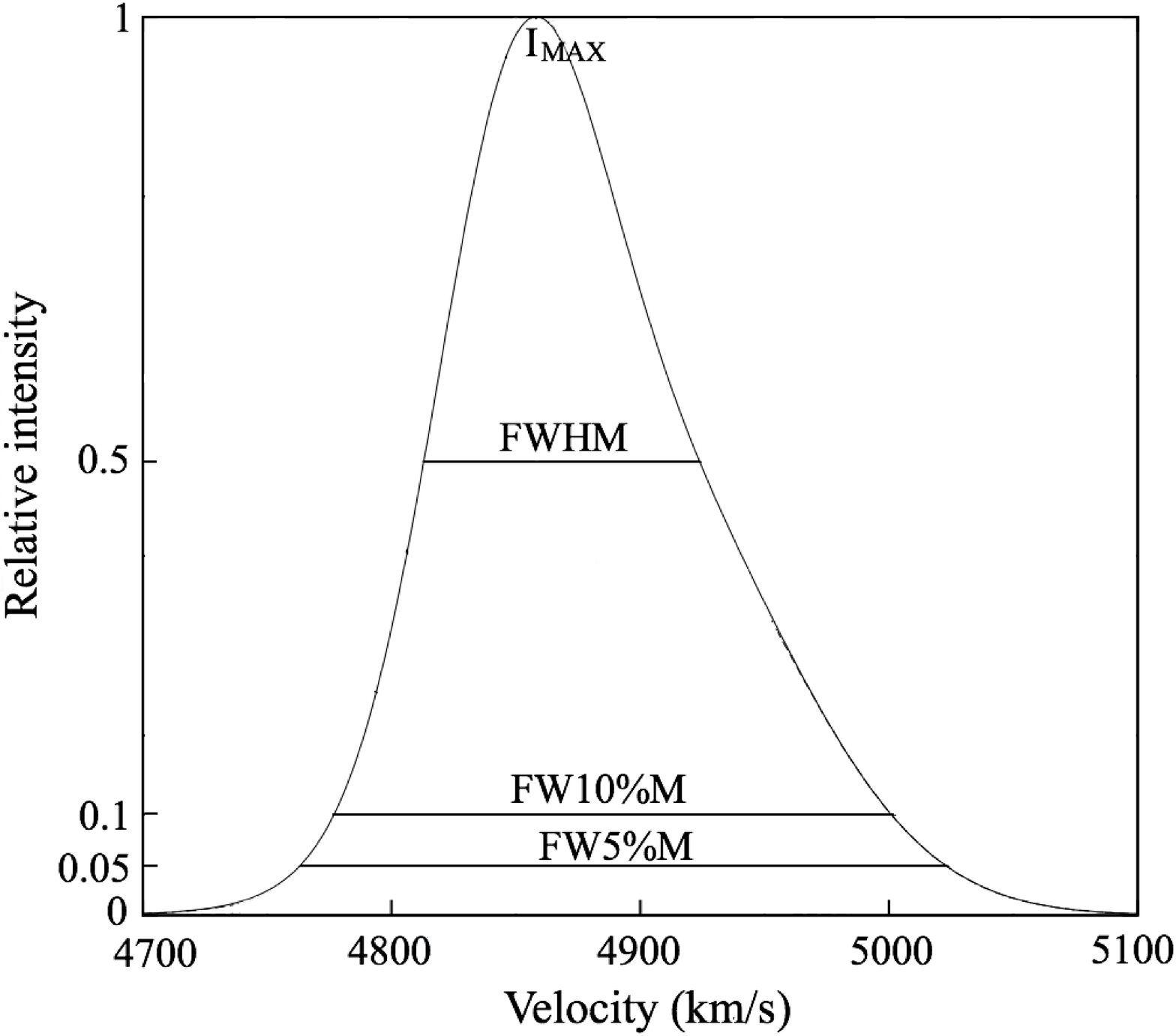}
 \includegraphics[width=0.45\textwidth]{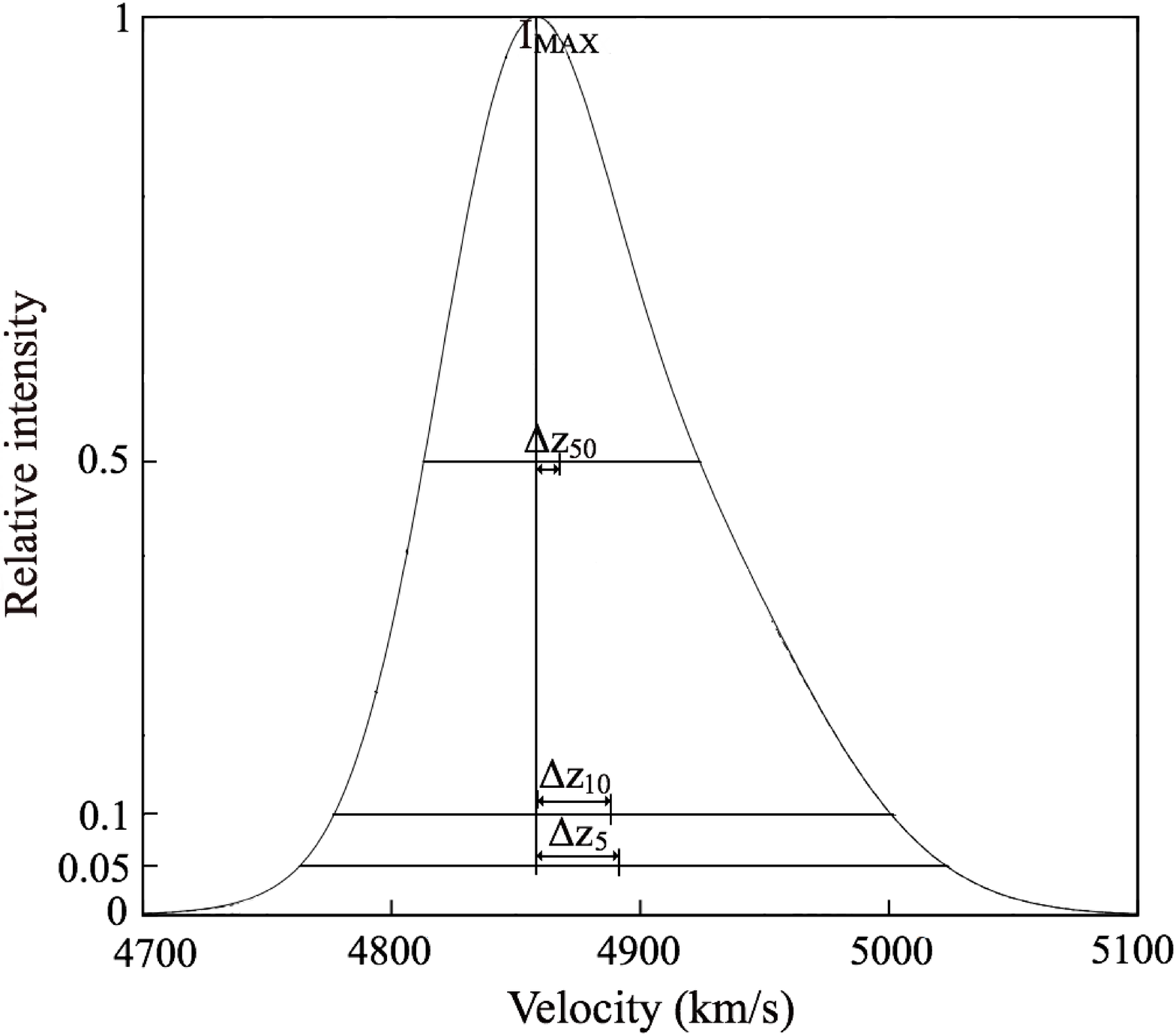}
    \caption{ The example of the line width (top) and intrinsic redshift measurements (bottom) of the broad (VBLR+ILR) component of ${\mathrm H\beta}$. The widths are measured at 50\%, 10\% and 5\% of the maximum (FWHM, FW10\%M and FW5\%M). The intrinsic redshifts (${\Delta z}_{50}$, ${\Delta z}_{10}$ and ${\Delta z}_{5}$) are measured as a difference between the centroid shift and the broad component peak, 
   at 50\%, 10\% and 5\% of the line maximum.
  }
    \label{fig:Figure2}
\end{figure}

\subsection{Measuring the widths and the intrinsic line shifts}

In order to measure the line widths and intrinsic shifts of the broad ${\mathrm H\beta}$ and \ion{Mg}{2}, we extract only the broad component from the composite H$\beta$ and \ion{Mg}{2} lines.

  The fitting procedure enables us to remove all overlapping lines, such as optical \ion{Fe}{2}, \ion{He}{2}, [\ion{O}{3}] and ${\mathrm H\beta}$ NLR in the case of ${\mathrm H\beta}$, so the sum of the ILR and VBLR ${\mathrm H\beta}$ components represents the clear broad ${\mathrm H\beta}$ profile (see Figure \ref{fig:Figure1}, top).
In the case of \ion{Mg}{2}, after removing the UV \ion{Fe}{2}, the broad profile of \ion{Mg}{2} consists of two Gaussians (see Figure \ref{fig:Figure1}, bottom).
We measure the FWHM for both lines, as well as Full Width at 10\% of the maximum (FW10\%M) and the Full Width at 5\% of the maximum (FW5\%M). The example of the line width measurements at different levels of line intensity is shown in Figure \ref{fig:Figure2} (top).

The intrinsic shift of the line is measured at different levels of the line intensity (at 50\%, 10\% and 5\%), as a centroid shift with respect to the broad line peak, as it is shown in Figure \ref{fig:Figure2} (bottom).
The intrinsic shift measured at 50\% of the line maximal intensity ($I_{MAX}$) is then:

$$\Delta z_{50} = z_{50} - z_{I_{MAX}}, \eqno(4)$$

\noindent where $z_{I_{MAX}}$ is the shift of the broad line peak, i.e. the maximum of the line intensity relative to the central wavelength, $z_{50}$ is the same but for the centroid at 50\% of the $I_{MAX}$.
 Similarly, the intrinsic shifts at the 10\% and 5\% of the $I_{MAX}$ are given as:

$${\Delta z}_{10} = z_{10} - z_{I_{MAX}}, \eqno(5)$$ 
$${\Delta z}_{5} = z_{5} - z_{I_{MAX}}, \eqno(6)$$ 
 where $z_{10}$ and $z_{5}$ are the shifts of the centroid at 10\% (5\% ) of the $I_{MAX}$, relative to the broad line peak.

\section{Results}

\indent 

To check the virialization of the broad ${\mathrm H\beta}$ and \ion{Mg}{2} lines, we analyze the relationships between their widths and intrinsic shifts. These parameters are measured at different levels of the line intensity (at 50 \%, 10 \% and 5 \% of Imax) and compared within one line and between ${\mathrm H\beta}$ and \ion{Mg}{2}.

 For the ${\mathrm H\beta}$ line, the measured shifts in the whole sample of 287 AGN are within the range from -445 ${\mathrm {km\ s}^{-1}}$ to 1045 ${\mathrm {km\ s}^{-1}}$ at 50\% of the $I_{MAX}$, from -1151  ${\mathrm {km\ s}^{-1}}$ to 3179 ${\mathrm {km\ s}^{-1}}$ at 10\% of the $I_{MAX}$ and from  -1469  ${\mathrm {km\ s}^{-1}}$ to 3897 ${\mathrm {km\ s}^{-1}}$ at 5\% of the $I_{MAX}$. In case of the \ion{Mg}{2} line, these values are within the range from -206  ${\mathrm {km\ s}^{-1}}$ to 1410 ${\mathrm {km\ s}^{-1}}$ at 50\% of the $I_{MAX}$, from -2771  ${\mathrm {km\ s}^{-1}}$ to 3678 ${\mathrm {km\ s}^{-1}}$ at 10\% of the $I_{MAX}$ and from -5224  ${\mathrm {km\ s}^{-1}}$ 
to 3678 ${\mathrm {km\ s}^{-1}}$ at 5\% of the $I_{MAX}$.

\bigskip

\indent We find that for the broad ${\mathrm H\beta}$ line, the line width is well correlated with the line intrinsic shift, measured at all intensity levels. For 50 \% of line intensity, the correlation between the 
FWHM and $\Delta z_{50}$ is $\rho$  $\approx$ 0.70, P $<$ 0.00001. Significant correlations are found also for the 10\% and 5\% of line intensity (FW10\%M vs. $\Delta z_{10}$, $\rho$  $\approx$  0.60, P $<$ 0.00001, FW5\%M vs. $\Delta z_{5}$, $\rho$ = 0.58, P $<$ 0.00001). The correlations are shown in Figure \ref{fig:Figure3}. The correlation coefficients between the widths and redshifts for the ${\mathrm H\beta}$ and \ion{Mg}{2} lines are listed in Table  \ref{table:Table1}.

\indent In the case of the \ion{Mg}{2} line, a correlation between the \ion{Mg}{2} width and line intrinsic shift is present only for 50 \% of the line intensity with $\rho$ = 0.59, P $<$ 0.00001.
However, it is interesting that for the 10 \% and 5 \% of the \ion{Mg}{2} intensity, the correlation between the \ion{Mg}{2} width and line intrinsic shift becomes opposite (see Figure \ref{fig:Figure4}), with $\rho$ = - 0.62, P $<$ 0.00001 for 10 \% of the line intensity, and $\rho$ = - 0.68, P $<$ 0.00001 for 5 \% of the line intensity.

\indent We analyze the relationships between the widths of the ${\mathrm H\beta}$ and \ion{Mg}{2} lines, measured at different levels of the line intensity. The results are shown in Figure \ref{fig:Figure5} and Table  \ref{table:Table2}. While there is a good correlation between the 
FWHMs of these lines ($\rho$ = 0.68, P $<$ 0.00001), there are no correlations between their FW10\%Ms and FW5\%Ms. Similar trend is seen for the line shifts: there is a correlation between the shifts of the lines at 50 \% of the intensity ($\Delta z_{50}$ ${\mathrm H\beta}$ vs. $\Delta z_{50}$ \ion{Mg}{2}, $\rho$ = 0.42, P $<$ 0.00001),
while there are no correlations between the $\Delta z_{10}$ (or $\Delta z_5$)  of ${\mathrm H\beta}$ and \ion{Mg}{2} (see Figure \ref{fig:Figure6} and Table  \ref{table:Table2}).

The measured values of the FWHMs, FW10\%Ms, FW5\%Ms and corresponding shifts of ${\mathrm H\beta}$ and \ion{Mg}{2}, for the complete sample of 287 AGN, are given in Tables  \ref{table:Table3} and  \ref{table:Table4}, in Appendix.

\indent  It is hard to estimate  the errors in the measured shifts and widths, since both, fitting procedure and S/N of a spectrum, are contributing. Since continuum subtraction probably has a large influence on the error, we make a test with an underestimate and overestimate of the underling continuum level for 3\% in the complete sample, and repeat the fitting procedure for ${\mathrm H\beta}$. We find that for ${\Delta z}_{50}$ the mean discrepancy with previous measurement is $\sim 20 {\mathrm {km\ s}^{-1}} \pm 100 {\mathrm {km\ s}^{-1}}$, while for the FWHM it is $\sim 200 {\mathrm {km\ s}^{-1}} \pm 390 {\mathrm {km\ s}^{-1}}$. For the widths and shifts measured at 10\% and 5\% of the $I_{MAX}$, the errors are larger. Nevertheless, the correlations between the shifts and widths found for the  10\% and 5\% of $I_{MAX}$ in the case of ${\mathrm H\beta}$  remain consistent. For FW10\%M vs. $\Delta z_{10}$ they are $\rho$ = 0.58,  P $<$ 0.00001 and $\rho$ = 0.55,  P $<$ 0.00001, for 3\% higher and 3\% lower continuum level respectively. In the case of FW5\%M vs. $\Delta z_{5}$, the correlations are: $\rho$ = 0.49,  P $<$ 0.00001 and $\rho$ = 0.52,  P $<$ 0.00001, respectively.

\subsection{The functional dependence between widths and intrinsic shifts}

We investigate the possible functional dependence of the ${\mathrm H\beta}$ and \ion{Mg}{2} line widths and intrinsic shifts. If the line widths and the intrinsic shifts are dominated by the gravitation of the BH, we expect that these parameters are linked with  a function of the form  $Y=m \cdot X^n$ (see Section 2, equation 3).

Therefore, we fit widths vs. shifts of ${\mathrm H\beta}$ and \ion{Mg}{2}, with  $Y=m \cdot X^n$, where $X = FWHMs$, and $Y = \Delta z_{50}$. The same is done for the widths and shifts measured at 10\% and 5\% of the maximal intensity.
The results are presented in Table  \ref{table:Table1} and Figure \ref{fig:Figure7}.

In the case of the ${\mathrm H\beta}$ line, the fit gives the exponent which is in agreement with the theoretical value of  $n = 2$, within the error-bars (see Table  \ref{table:Table1}). The agreement is slightly better for the widths and shifts measured at 50\% of Imax ($n = 1.85\pm0.19$).
 However, in the case of  \ion{Mg}{2}, the exponent which is in  agreement with the theoretical value is obtained only for the widths and shifts measured at 50\% of Imax
($n = 2.06\pm0.27$), while for the 10\%  and 5\% of Imax, the exponent is $n < 0$ (see Table  \ref{table:Table1}). 

In order to find more accurate relationships between widths and shifts of ${\mathrm H\beta}$ and \ion{Mg}{2}, we perform the same fitting for  the subsample of 123 AGN, where both lines, ${\mathrm H\beta}$ and \ion{Mg}{2}, are redshifted (see Sec 3.1).
The results are shown in Figure \ref{fig:Figure8}, and in Table  \ref{table:Table1}. 

\begin{figure}[h!]
    \centering
        \includegraphics[width=0.42\textwidth]{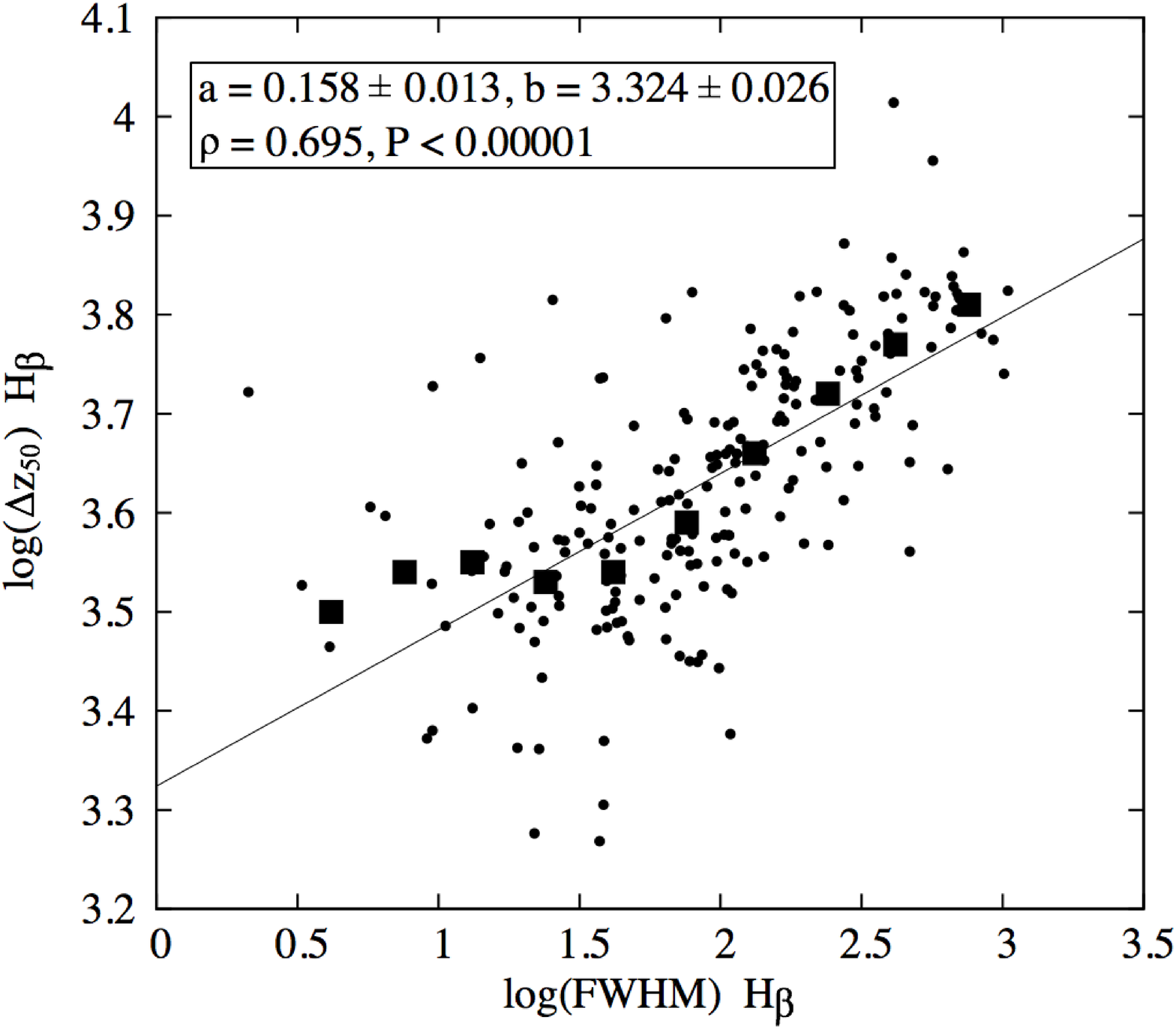}
        \includegraphics[width=0.42\textwidth]{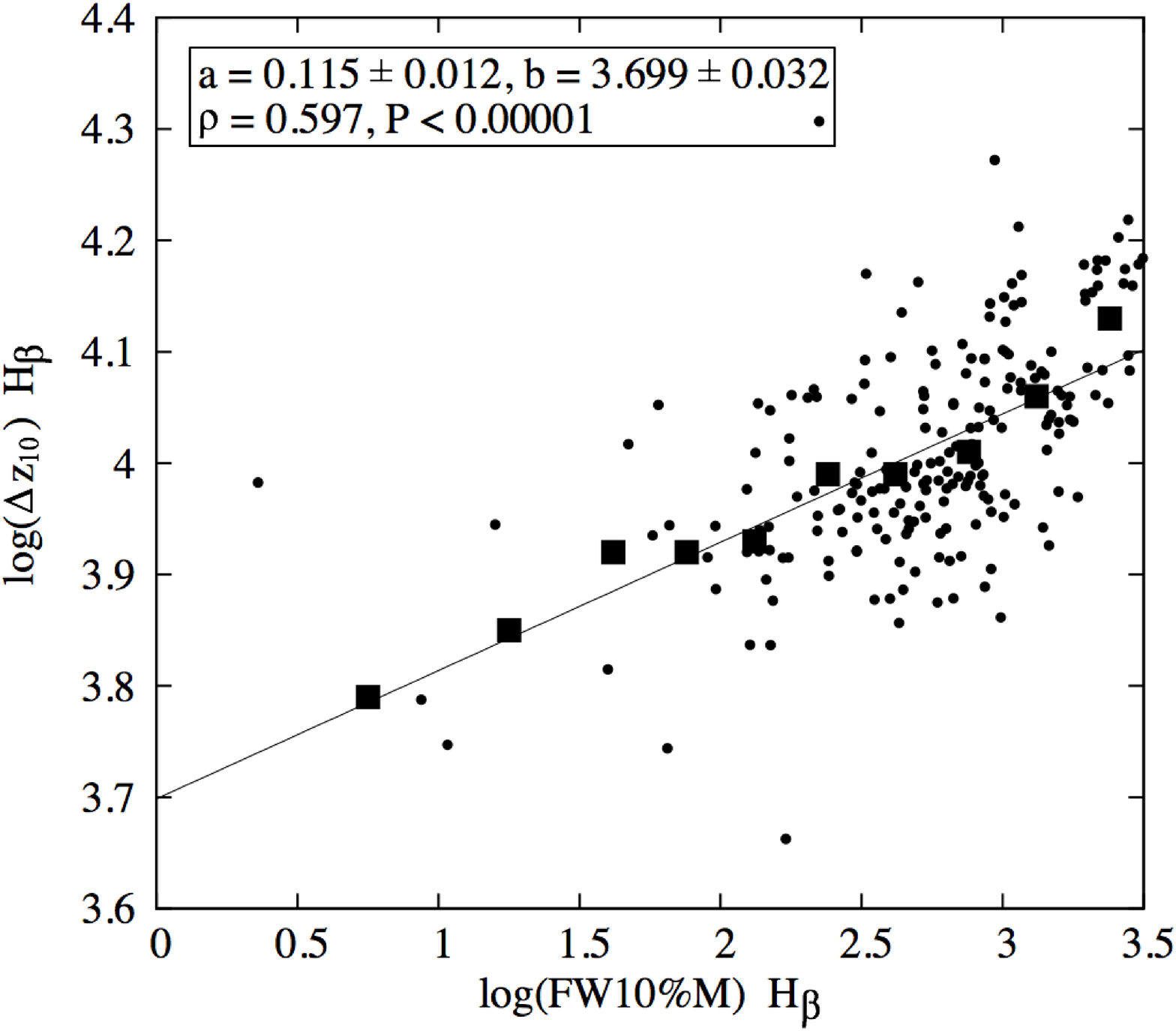}
        \includegraphics[width=0.42\textwidth]{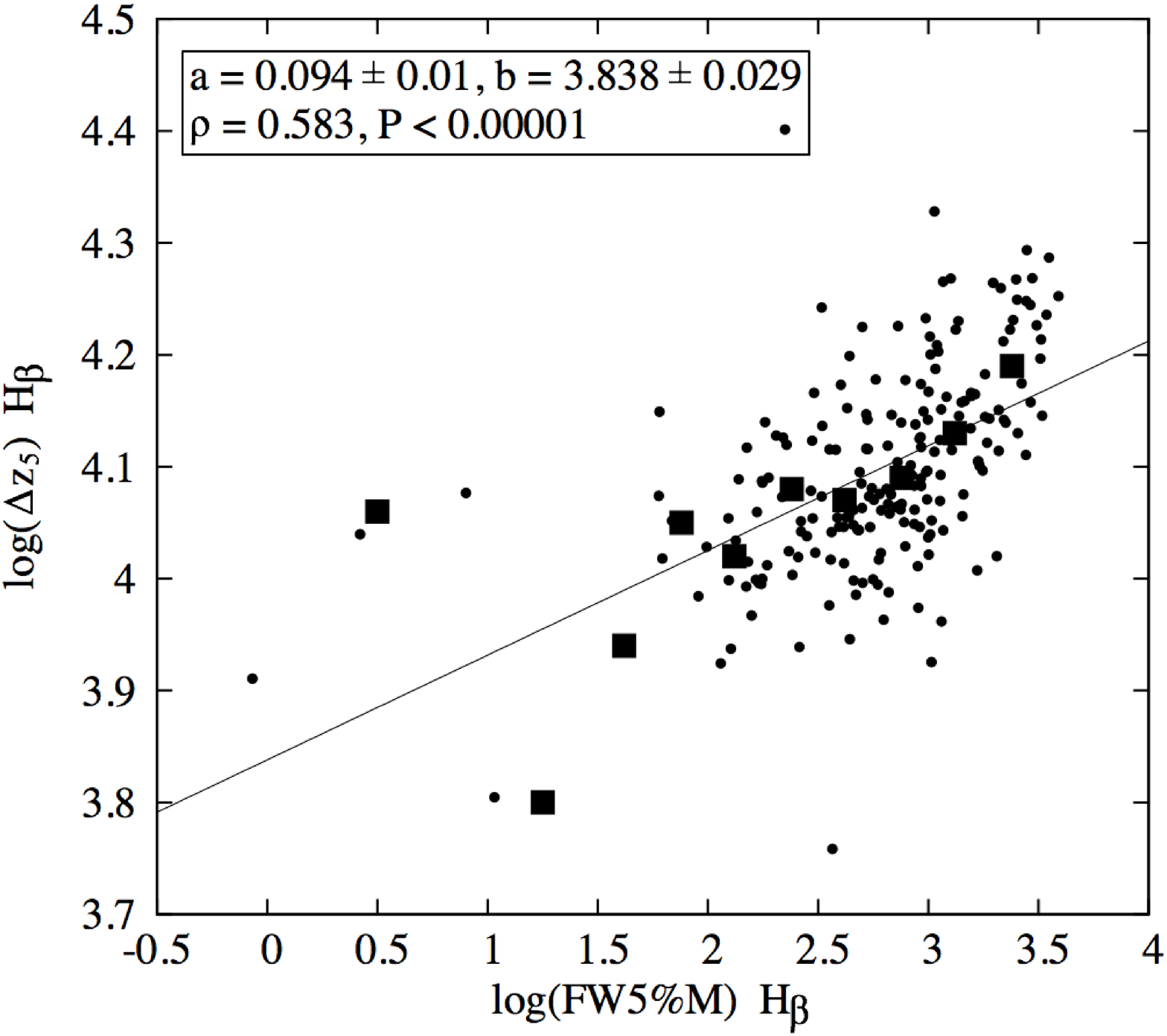}
    \caption{Correlation between the widths and intrinsic shifts of the broad ${\mathrm H\beta}$ lines (given in km/s), measured for 50 \% Imax (top), 10 \%  Imax (middle), and 5 \% of the Imax (bottom). Data are fitted with a linear function: $Y=a+b \cdot X$. The coefficients a and b, as well as Spearman coefficient of correlation ($\rho$) and P-value are shown on plots. The binned values are assigned with black squares.}
    \label{fig:Figure3}
\end{figure}

\begin{figure}[h!]
    \centering
        \includegraphics[width=0.47\textwidth]{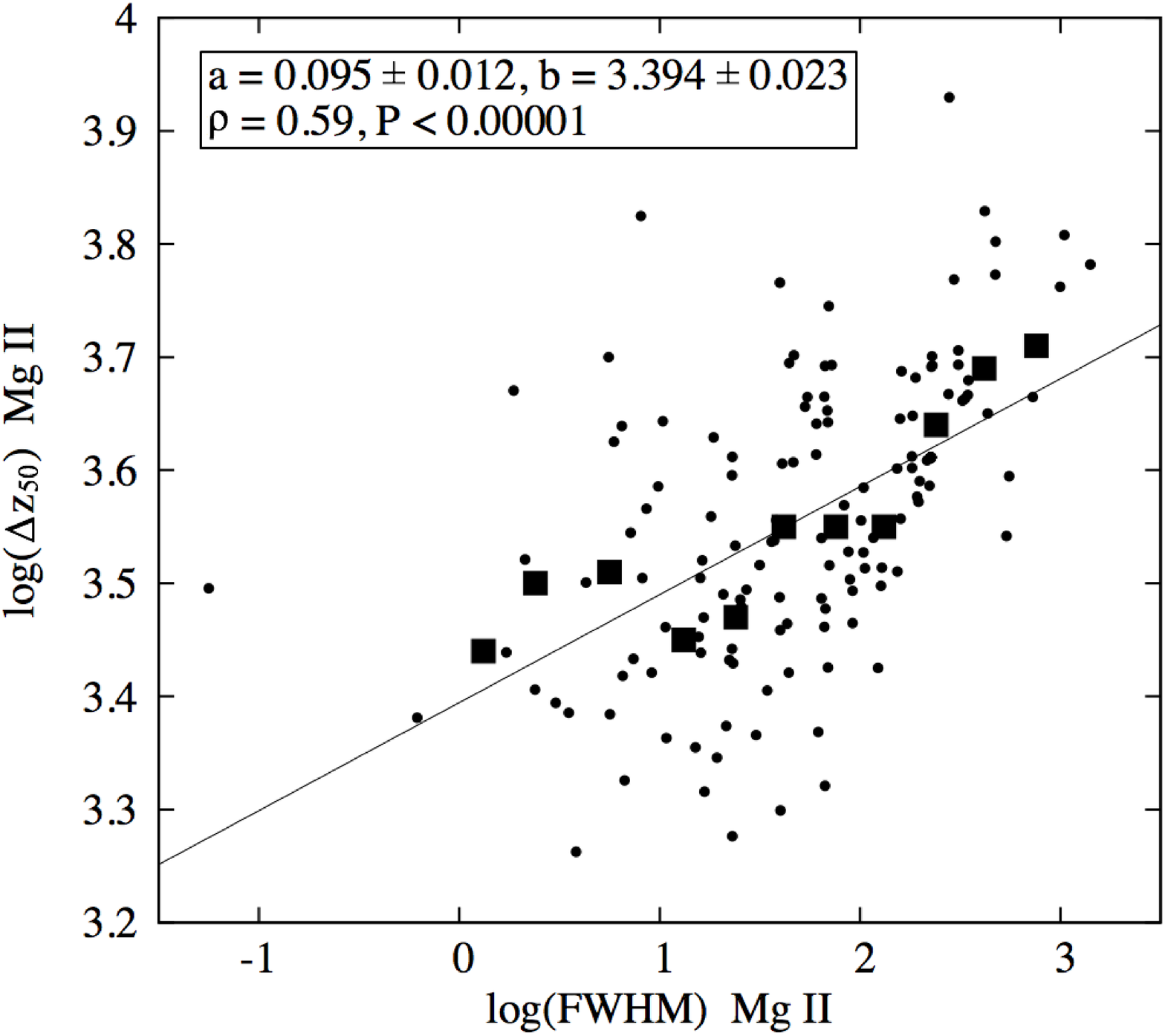}
        \includegraphics[width=0.47\textwidth]{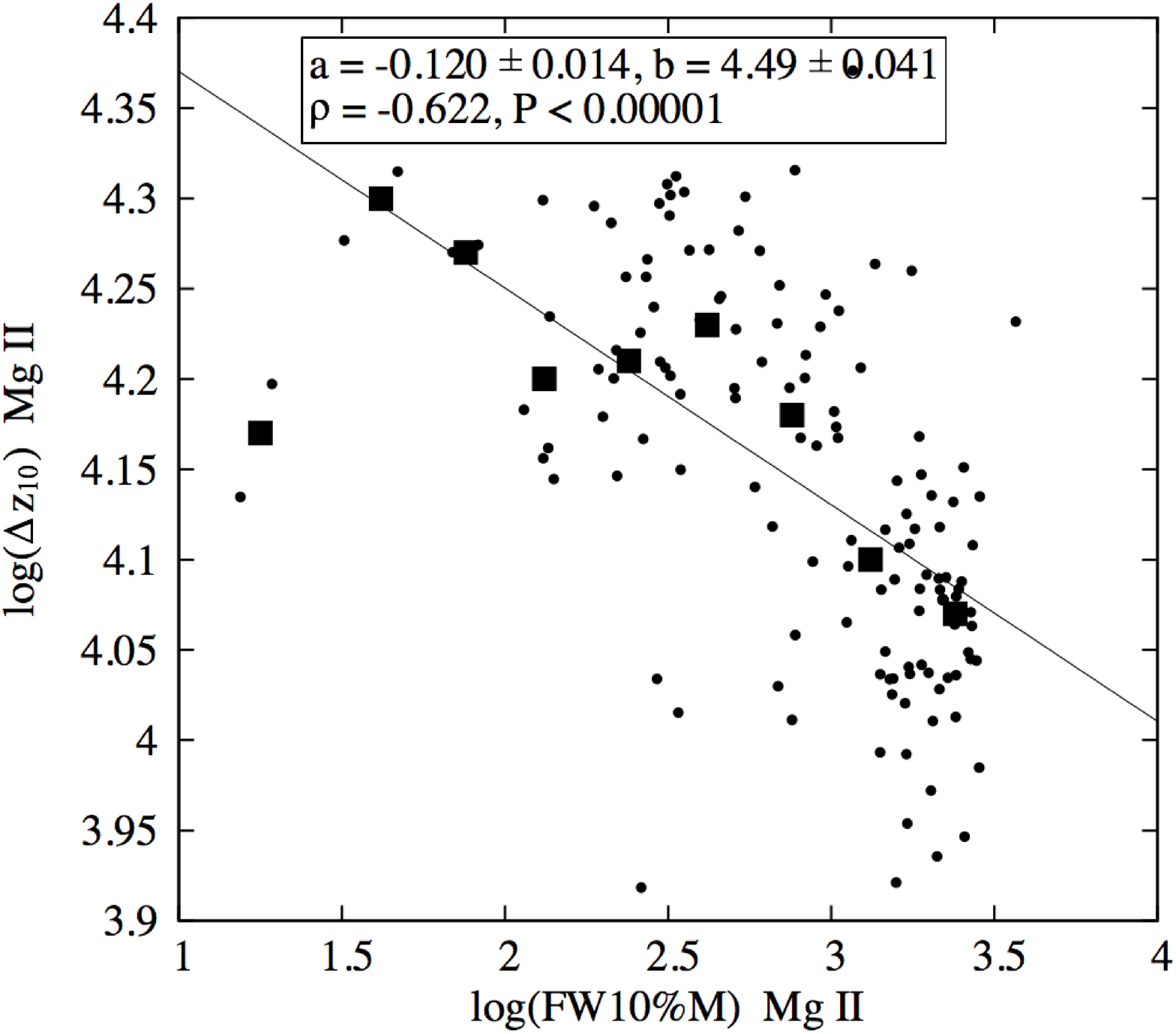}
        \includegraphics[width=0.47\textwidth]{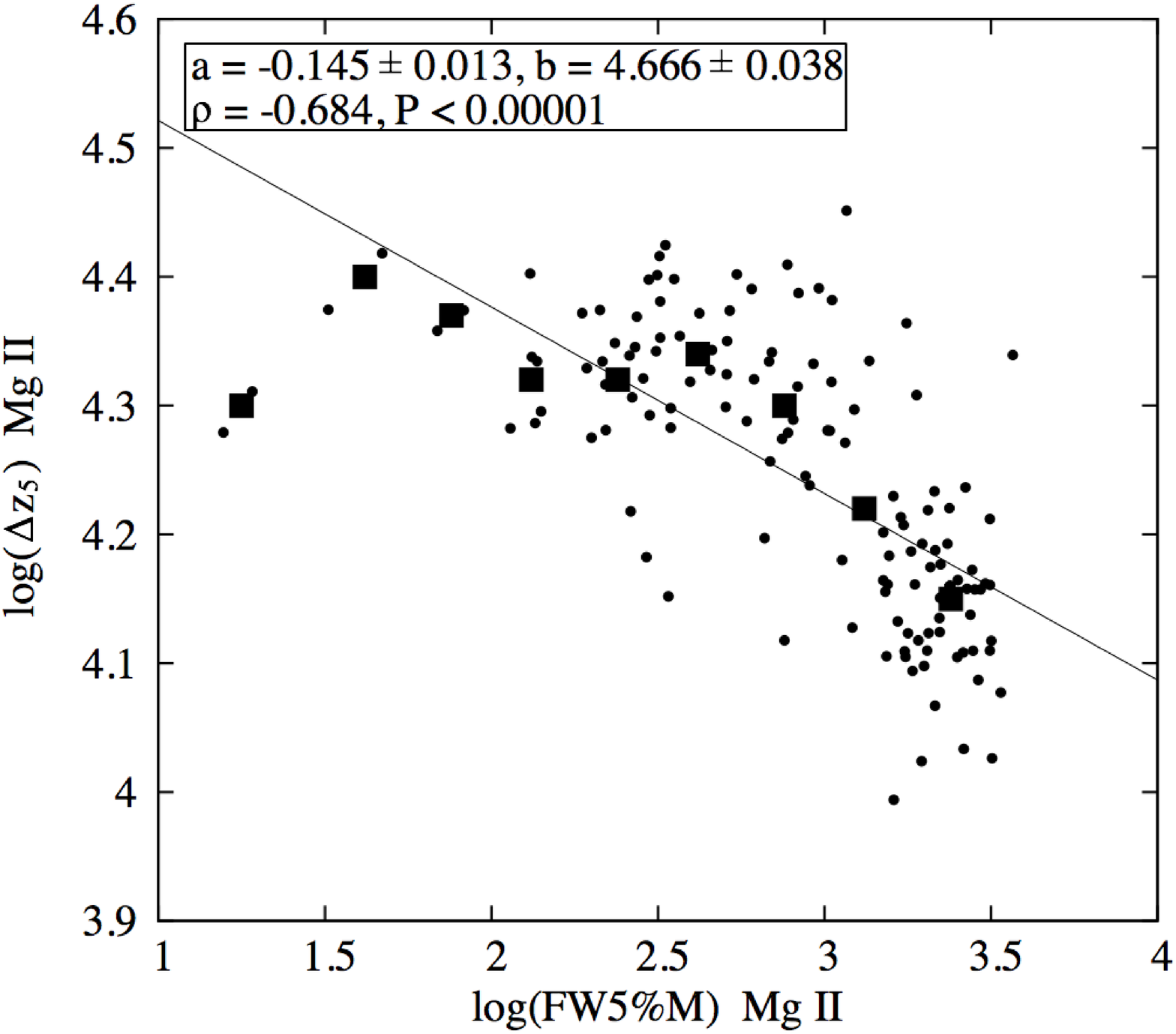}

    \caption{ The same as in Figure \ref{fig:Figure3} but for \ion{Mg}{2}.}
    \label{fig:Figure4}
\end{figure}

\section{Discussion}

\indent Since the broad ${\mathrm H\beta}$ and \ion{Mg}{2} lines are widely used as virial estimators of AGN BH masses, it is important to investigate their applicability for that purpose, and to check  
whether the intrinsic redshifts of these lines represent gravitational redshifts, i.e. could they be used as good virial estimators.

 There is the evident difference in the profiles of the \ion{Mg}{2} and ${\mathrm H\beta}$ lines. \citet{Kovacevic2015} compared the average widths of Gaussians which fit the core and the wings of the \ion{Mg}{2} and ${\mathrm H\beta}$, and found that the core of \ion{Mg}{2} is slightly narrower than the core of ${\mathrm H\beta}$, while wings of the \ion{Mg}{2} are significantly broader than the ${\mathrm H\beta}$ ones. Furthermore, there is a correlation between the \ion{Mg}{2} and ${\mathrm H\beta}$ core widths, as well as between their shifts, while there is no correlation between the kinematical properties of their wings \citep{Kovacevic2015}. This implies that their cores probably originate from the kinematicaly connected regions, while this can not be stated for their wings. In our sample of 287 objects, there are only 123 AGN which have no blueshift in ${\mathrm H\beta}$ and \ion{Mg}{2}. Some objects have a blueshift in one of these two lines, which means that if one of these two lines is a good 
virial estimator for one object, the other line can be strongly influenced by some other effects beside the gravitational, and could not be taken as a good virial estimator. This implies as well that emission in wings of these lines probably originate from kinematicaly different emission region.

Considering only the AGN subsamples with redshifted ${\mathrm H\beta}$ or \ion{Mg}{2}, we find expected linear correlations between logarithms of the FWHMs and $\Delta z_{50}$ for both lines, which implies that the intrinsic redshifts are indeed connected with the gravitational redshift of these lines. However, while correlation between FWHM and intrinsic redshift is significant at all levels of Imax for ${\mathrm H\beta}$, for \ion{Mg}{2} the anti-correlation becomes significant for 10\% and 5\% of Imax. 

 Similarly, the widths  of  these two lines are well correlated at 50\% of the line intensity, as well as the intrinsic shifts, while there are no correlations between these parameters  at 10\% and 5\% of Imax.

The above result is confirmed with fitting of the functional dependence between the widths and shifts of these lines, as well.  The theoretically predicted dependence  $\Delta z_{50} = const \cdot FWHM^2$ (see Sec 2) is obtained from the best fit  with the function $Y=m \cdot X^n$. The estimated $n$ values are in agreements with predicted $n = 2$, within the error-bars for both lines, ${\mathrm H\beta}$ and \ion{Mg}{2}. However, while for ${\mathrm H\beta}$ the mentioned relationship is valid also for widths and intrinsic redshifts at 10\% and 5\% of Imax, for \ion{Mg}{2} wings our initial assumption of the virilized gas failed. The relationships between the \ion{Mg}{2} widths and intrinsic shifts at 10\% an 5\% of Imax do not match the theoretical prediction.

 Numerous UV \ion{Fe}{2} lines form broad features which is hard to be distinguished from the \ion{Mg}{2} wings. Therefore, one can suspect that the \ion{Mg}{2} wings shape can be affected by subtraction of the Balmer continuum and UV \ion{Fe}{2} lines which overlap with \ion{Mg}{2}. Surely, this introduces an uncertainty for FWHM and $\Delta z$ measurements close to continuum level, at 10\% and 5\% of Imax. However, we do not expect that this uncertainty produces significant anti-correlations between widths and intrinsic shifts in the \ion{Mg}{2} wings, which are detected ($\rho$ = - 0.62 and $\rho$ = - 0.68, P $<$ 0.0001, for 10\% and 5\% of Imax).

\indent The blueshifts in both lines, and not applicable virial assumption in case of \ion{Mg}{2} line wings, can be explained by outflow of the BLR gas, caused by the radiation pressure force or some
other unknown mechanisms \citep{Ilic2010}.

\section{Conclusions}

\indent In this paper we investigate the virilization of the broad ${\mathrm H\beta}$ and \ion{Mg}{2} lines in the sample of $287$ type 1 AGN taken from the SDSS database. 
The aim of this work is to test whether the broad ${\mathrm H\beta}$ and \ion{Mg}{2} are indeed good virial estimators for all objects, and to check if the intrinsic redshift in these lines is connected to the gravitation redshifts, i.e.\ can be used for the BH mass estimation.
For that purpose, we measure the widths and intrinsic shifts of those lines, at 50\%, 10\% and 5\% of the maximal intensity.\ We analyze the correlations and functional dependences between centroid redshifts and corresponding widths of those lines, and compare it with theoretically predicted relationships.

\bigskip

\indent From this research we can conclude that:

\begin{enumerate}

\item 
 For the AGN sample with redshifted ${\mathrm H\beta}$ line, there are good correlations between all measured shifts and widths (at 50\%, 10\% and 5\% of Imax).\  Also, the theoretically expected relationship $\Delta z_{50}\sim FWHM^2$ is confirmed, within the error-bars. The same relationship is confirmed for widths and corresponding intrinsic redshifts, measured at different intensity levels (10\% and 5\% of Imax).
This implies that ${\mathrm H\beta}$ is a good virial estimator of AGN BH masses, and that the intrinsic redshift of the ${\mathrm H\beta}$ line is dominantly caused by gravitational effects. However,
there is a group of AGN with a blue asymmetry in the broad emission lines. Therefore, to use the ${\mathrm H\beta}$ line one should check the asymmetry of ${\mathrm H\beta}$, and in the case of the red asymmetry, it probably can be used for the BH mass estimation, while in the case of the blue asymmetry, it should be taken with caution.

\item
For the AGN sample with the intrinsic redshift in the broad \ion{Mg}{2} line, there is the correlation between intrinsic redshift and width but only at 50\% of Imax. The theoretically expected relationship $\Delta  z_{50}\sim FWHM^2$ is confirmed as well, but it failed for widths and intrinsic redshifts at 10\% and 5\% of Imax. As the widths and shifts are measured closer to the continuum level in the wings of the \ion{Mg}{2} lines, an anti-correlation between 
widths and corresponding redshifts begins to appear. Consequently, the \ion{Mg}{2} line can be used as a virial estimator for AGN with the red asymmetry, but only at 50\% of Imax. The intrinsic redshift at that level probably can be used for the BH mass estimation. On the other hand, the widths and shifts measured in wings of these lines can not be used for this purpose. 
The same as for ${\mathrm H\beta}$, in the case of the blue asymmetry, \ion{Mg}{2} should be taken with caution as the virial estimator.
\item
 This research open some new questions considering the anti-correlations between gravitational redshift and widths in the wings of \ion{Mg}{2} line. Also, the number of the \ion{Mg}{2} lines with the blue asymmetry is larger compared to ${\mathrm H\beta}$.
It seems that the broader component of \ion{Mg}{2} (which contribute to the line wings) is coming from the region where some other mechanisms, except the gravity cannot be neglected. 
 Candidates for this mechanism are gas outflows from the BLR probably caused by the radiation pressure force or accretion disk wind.

\end{enumerate}

\section*{Acknowledgments}

This investigation is supported by the Ministry of Education, Science and Technological Development of Republic of Serbia in the frame of Project 176001 \textit{Astrophysical spectroscopy of extragalactic objects}. We thank to an anonymous referee for
very useful comments.

\pagebreak
\
\begin{figure}[h!]
    \centering
        \includegraphics[width=0.43\textwidth]{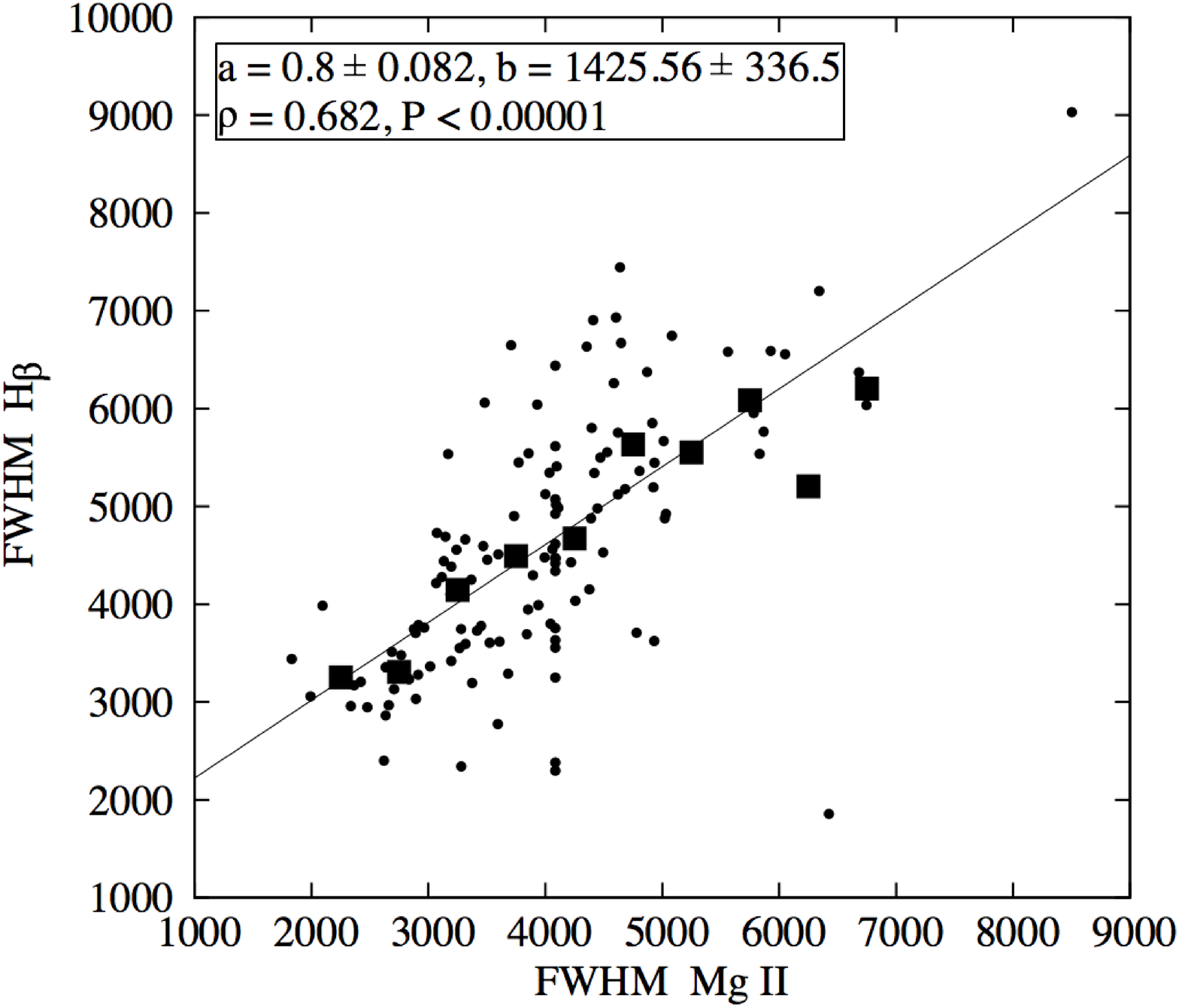}
        \includegraphics[width=0.43\textwidth]{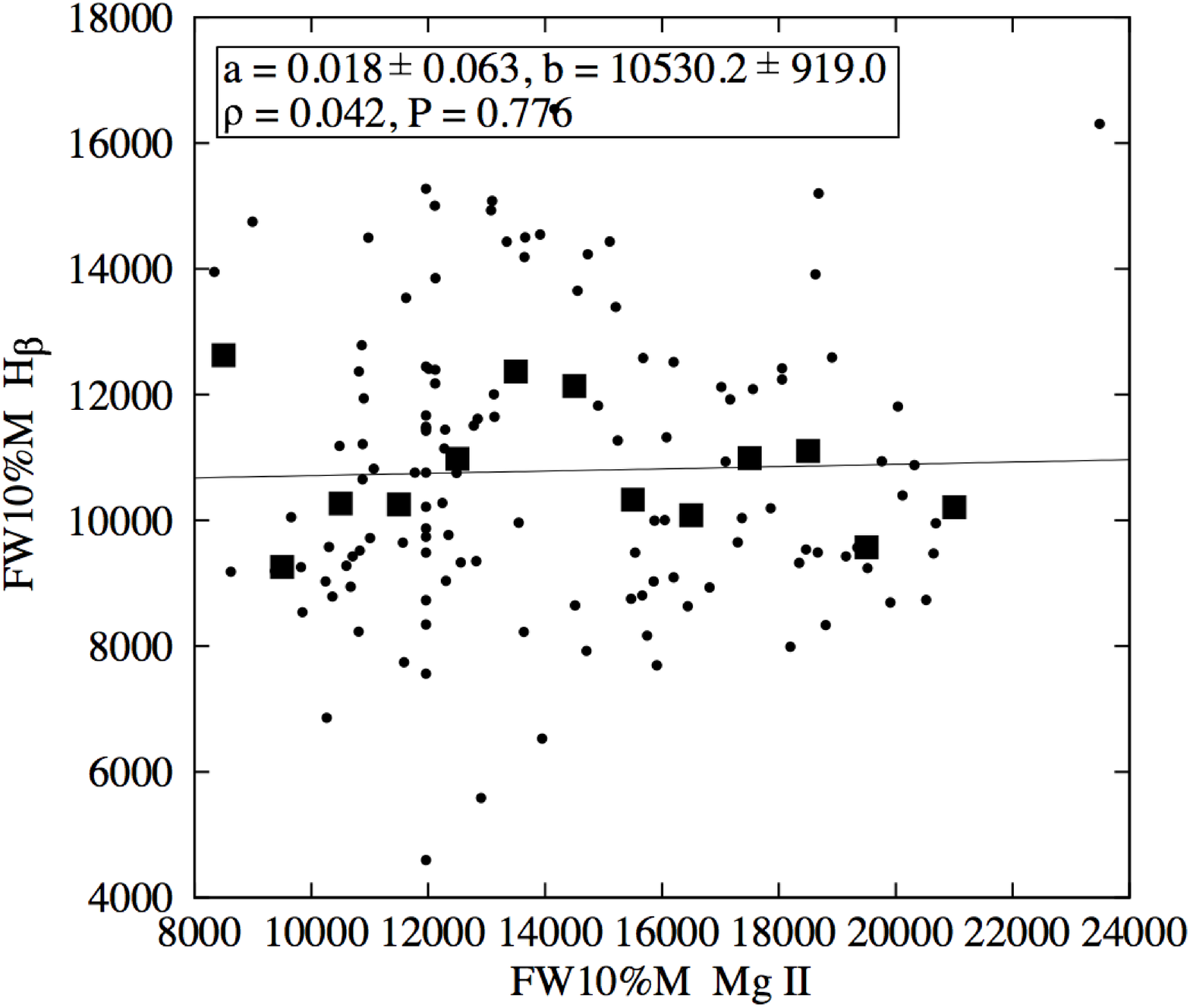}
        \includegraphics[width=0.43\textwidth]{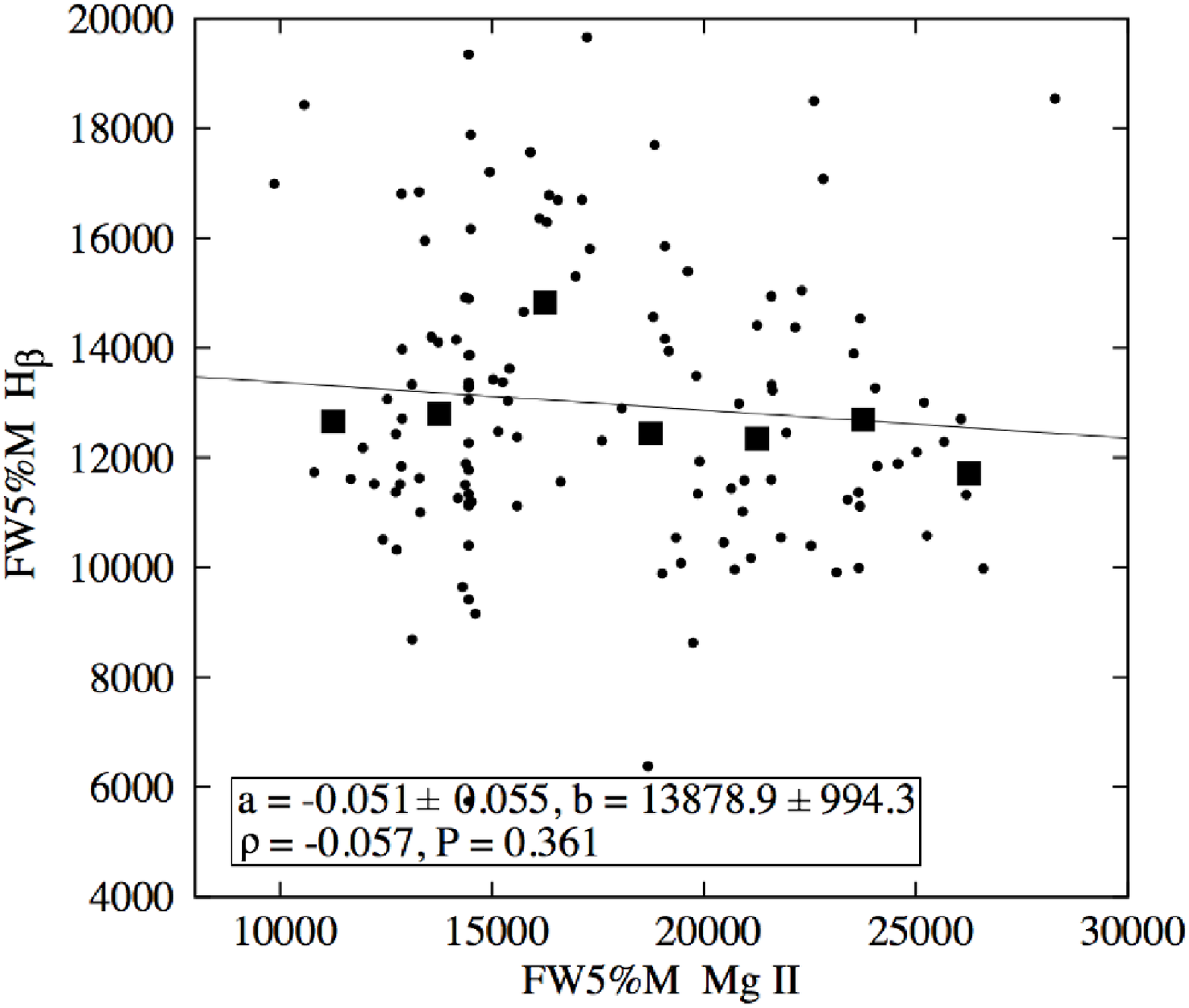}
    \caption{Top: the correlation between the FWHMs of the ${\mathrm H\beta}$ and \ion{Mg}{2} lines. Middle and bottom: no correlation between the FW10\%Ms and FW5\%Ms of these lines. The notation is the same as in Figure \ref{fig:Figure3}. }
    \label{fig:Figure5}
\end{figure}

\begin{figure}[h!]
    \centering
        \includegraphics[width=0.43\textwidth]{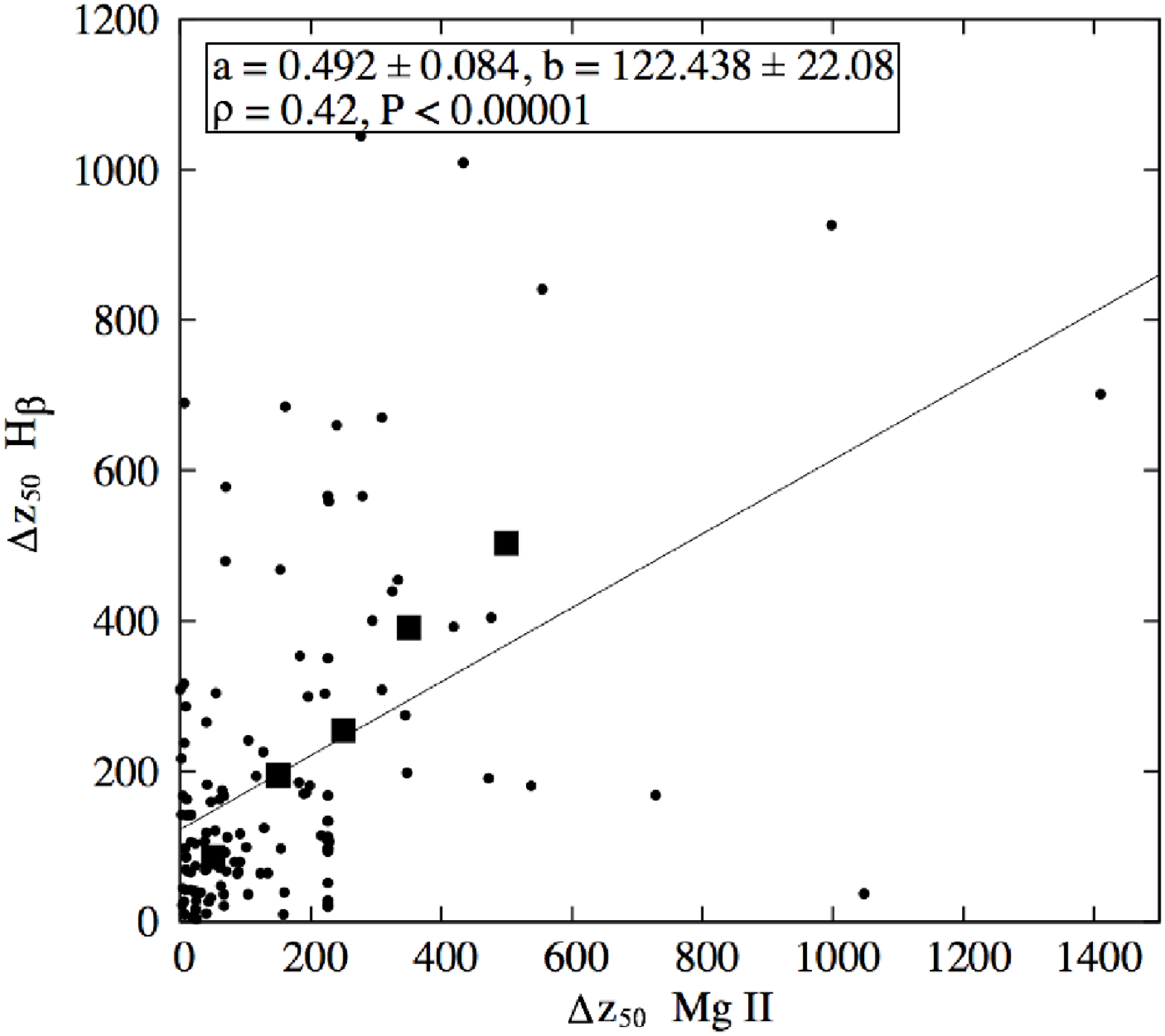}
        \includegraphics[width=0.43\textwidth]{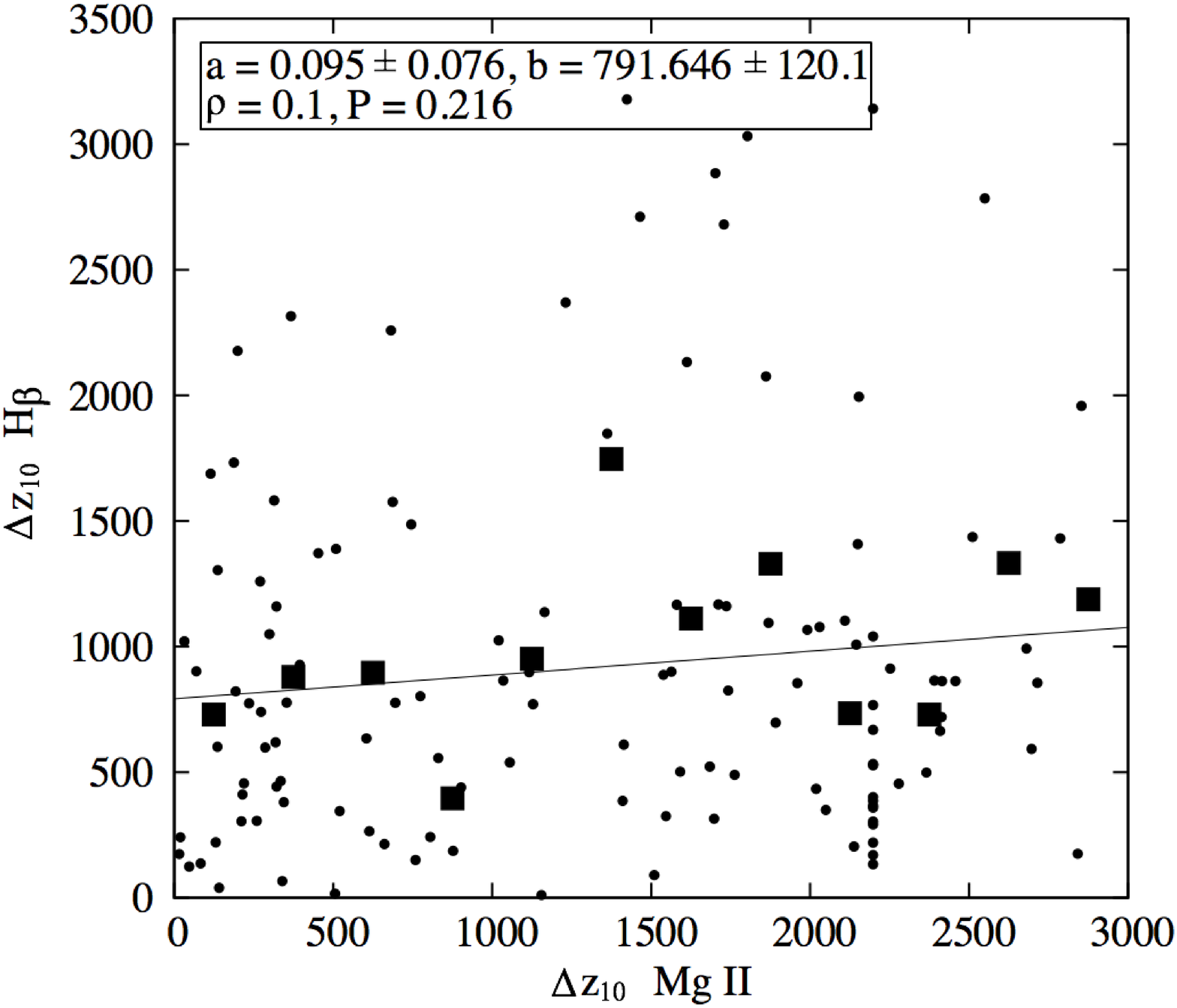}
        \includegraphics[width=0.43\textwidth]{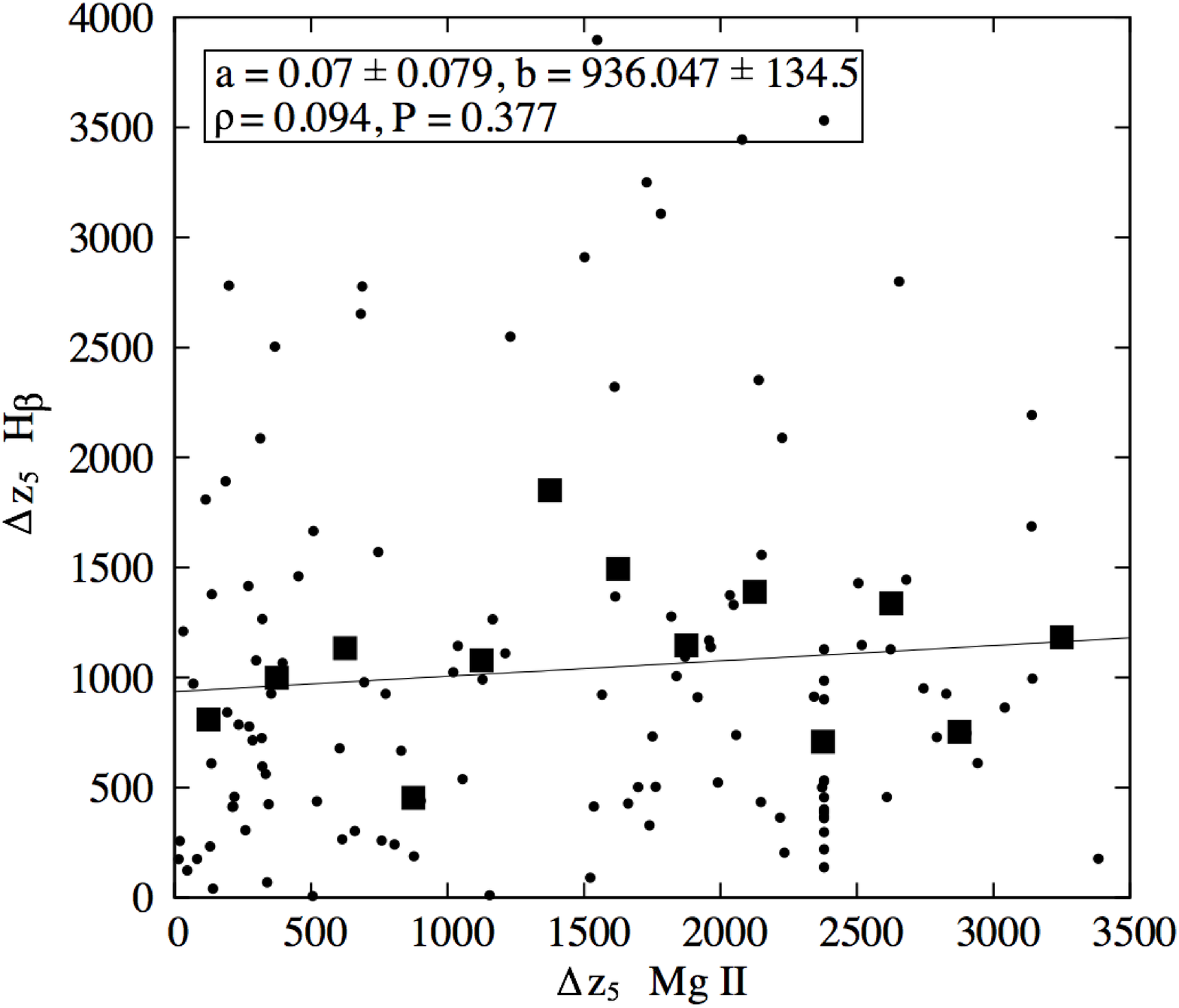}
    \caption{Top: the correlation between the redshifts ${\Delta z}_{50}$ of the ${\mathrm H\beta}$ and \ion{Mg}{2} lines. Middle and bottom: no correlation between the redshifts ${\Delta z}_{10}$ and ${\Delta z}_{5}$ of these lines. The notation is the same as in Figure \ref{fig:Figure3}.}
    \label{fig:Figure6}
\end{figure}

\begin{figure}[h!]
    \centering
        \includegraphics[width=0.47\textwidth]{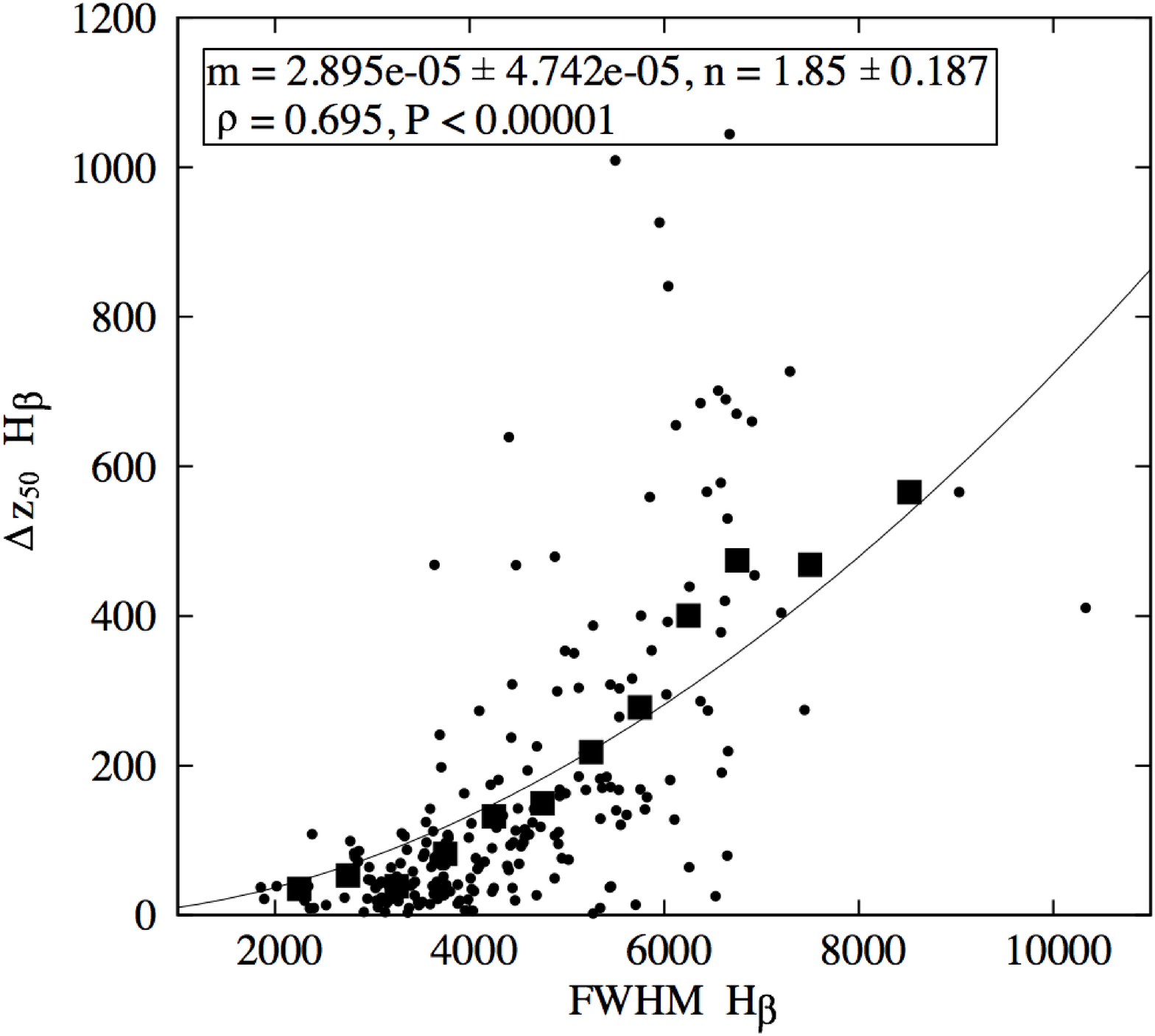}
        \includegraphics[width=0.47\textwidth]{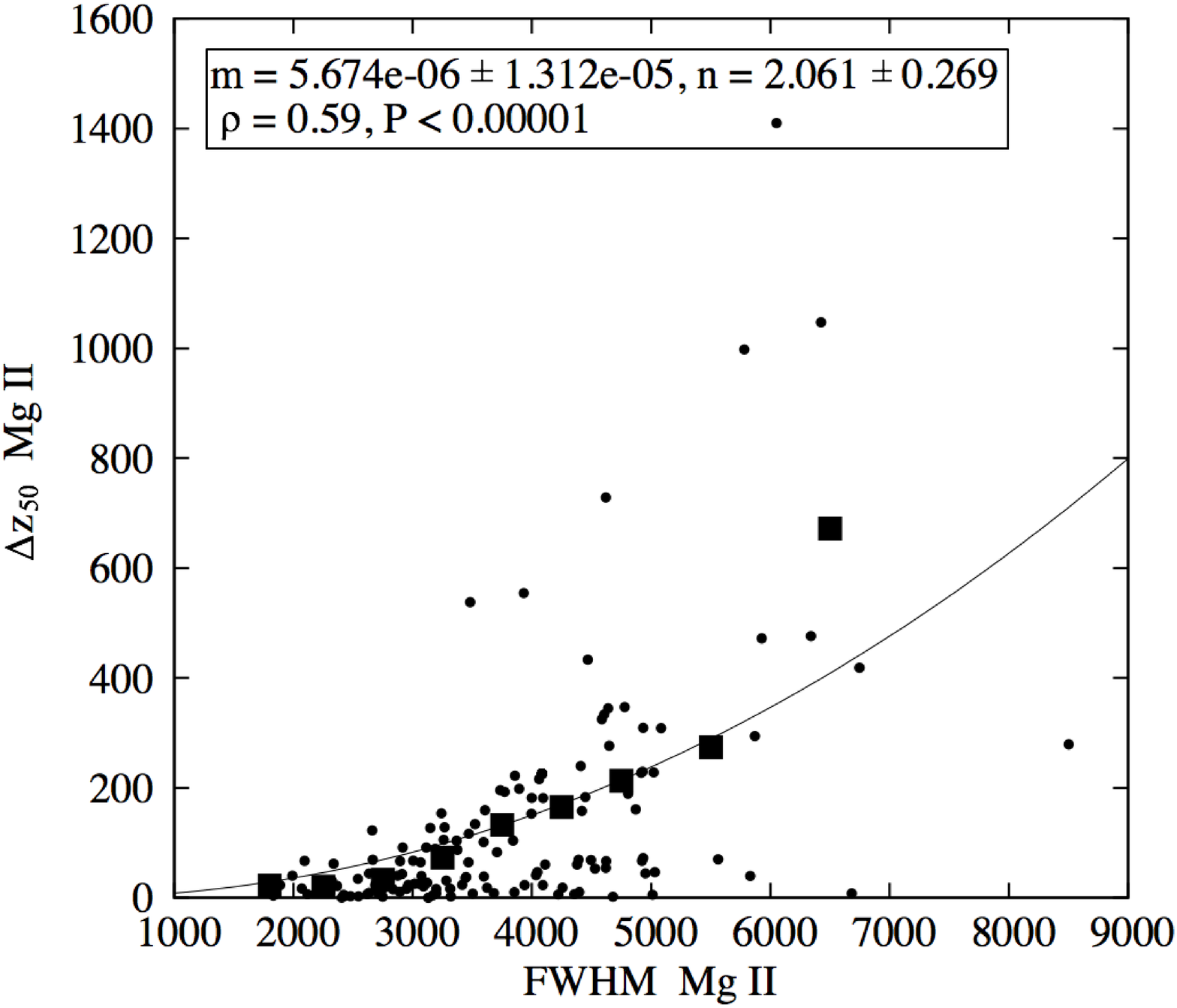}

    \caption{ Relationship between widths (km/s) and intrinsic shifts (km/s) of ${\mathrm H\beta}$ (top) and \ion{Mg}{2} (bottom). The observations are fitted with the function: $Y=m \cdot X^n$. Binned values are shown with black squares.} 
    \label{fig:Figure7}
\end{figure}

\begin{figure}[h!]
    \centering
        \includegraphics[width=0.47\textwidth]{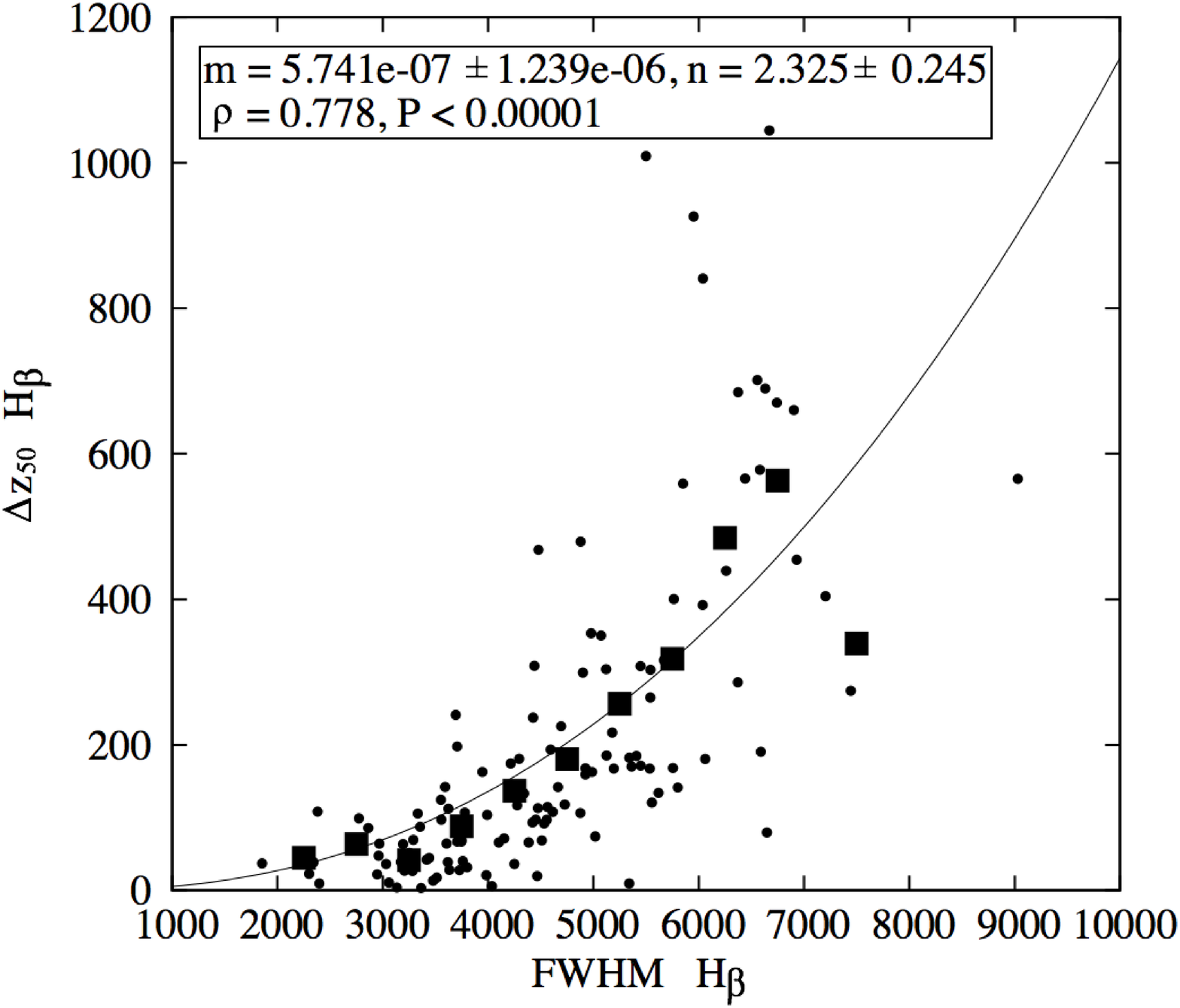}
        \includegraphics[width=0.47\textwidth]{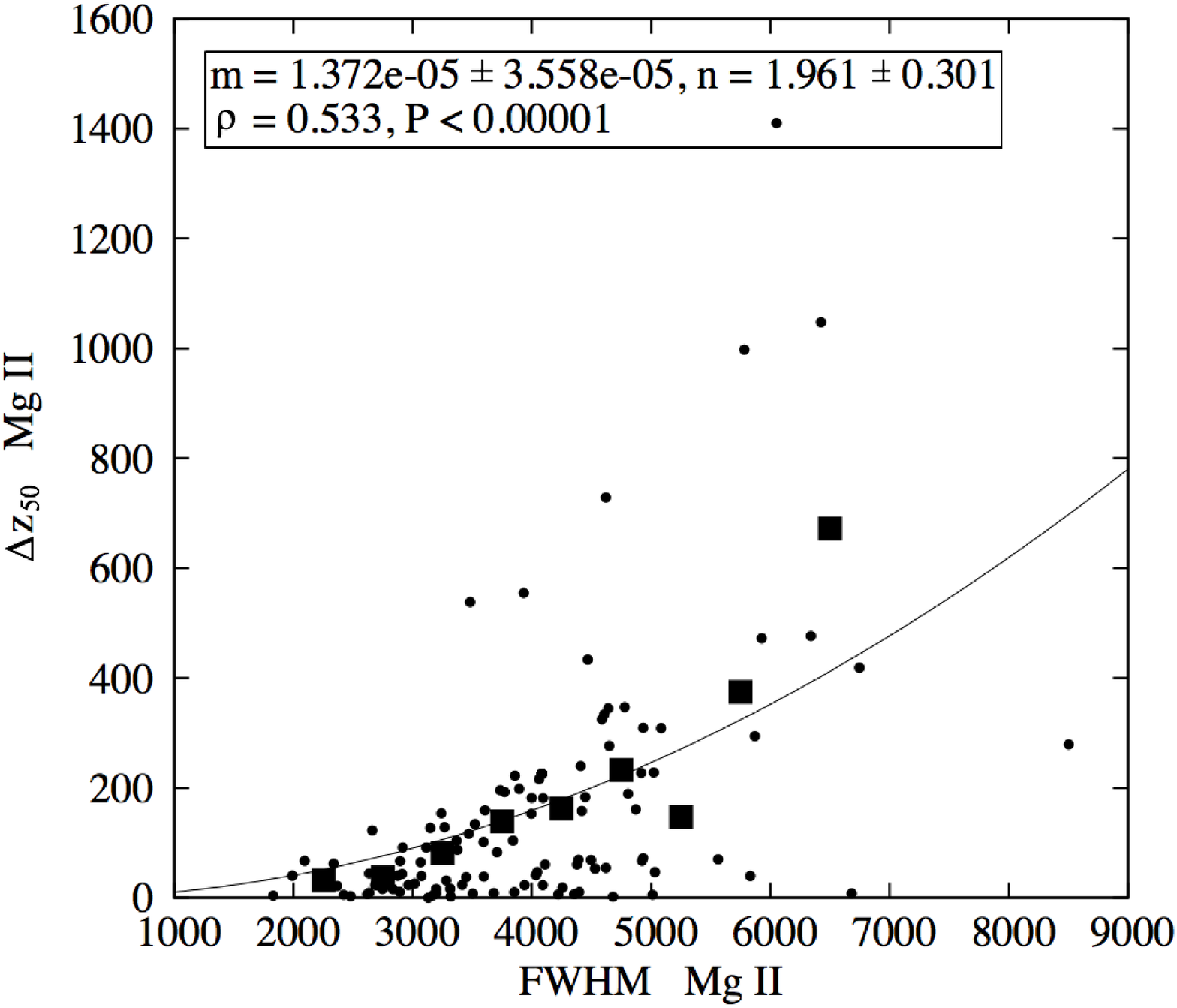}

    \caption{The same as in Figure \ref{fig:Figure7}, just for the subsample of 123 AGN, in which both, ${\mathrm H\beta}$ and \ion{Mg}{2} lines, are redshifted. All spectra with the blueshift in one of these two lines are rejected. } 
   
 \label{fig:Figure8}
\end{figure}

\clearpage

\begin{table*}[t]
  \centering

\caption{The widths (X) and  intrinsic shifts (Y) of the ${\mathrm H\beta}$ and Mg II are fitted with  function: $Y=m \cdot X^n$. The parameters $m$ and $n$, obtained from the fit, as well as the Spearman coefficients of correlation ($\rho$) and $P$-values, are also listed.}
 \label{table:Table1}

\begin{tabular}{|c|c|c|c|c|c|}
\hline

 widths vs. shifts & number of AGN & $m$ & $n$ &$\rho$&P\\
\hline

FWHM ${\mathrm H\beta}$ vs. ${\Delta z}_{50}$  ${\mathrm H\beta}$ &209  &2.90E-5$\pm$4.74E-5  &  1.85$\pm$0.19 & 0.69    & P$<$0.00001\\
FW10\%M ${\mathrm H\beta}$ vs. ${\Delta z}_{10}$ ${\mathrm H\beta}$&209  &6.40E-5$\pm$1.14E-5  &  1.77$\pm$0.19 & 0.60    & P$<$0.00001\\
FW5\%M ${\mathrm H\beta}$  vs. ${\Delta z}_{5}$ ${\mathrm H\beta}$& 209 &7.10E-5$\pm$1.39E-4  &  1.74$\pm$0.20 & 0.58  & P$<$0.00001\\
FWHM Mg II vs. ${\Delta z}_{50}$ Mg II &150 &  5.67E-6$\pm$1.31E-5  &2.06$\pm$0.27 & 0.59 & P$<$0.00001 \\
FW10\%M Mg II vs. ${\Delta z}_{10}$ Mg II& 150 & 1.26E+9$\pm$2.57E+9 & -1.45$\pm$0.22 & -0.62 & P$<$0.00001\\
FW5\%M Mg II vs. ${\Delta z}_{5}$ Mg II&150&  2.133E+8$\pm$3.98E+8 &-1.23$\pm$0.30 & -0.68 & P$<$0.00001\\

FWHM ${\mathrm H\beta}$ vs. ${\Delta z}_{50}$  ${\mathrm H\beta}$ &123  &5.74E-7$\pm$1.29E-6  &  2.33$\pm$0.25 & 0.78    & P$<$0.00001\\
FWHM Mg II vs. ${\Delta z}_{50}$ Mg II &  123  &1.37E-5$\pm$3.56E-5 &1.96$\pm$0.30 & 0.53 & P$<$0.00001 \\

\hline

\end{tabular}

\end{table*}

\begin{table*}[t]
  \centering
\caption{The intrinsic redshifts and widths of \ion{Mg}{2} (X) and ${\mathrm H\beta}$ lines (Y) are fitted with linear function: $Y=a \cdot X + b$. The parameters $a$ and $b$, obtained from the fit, as well as the Spearman coefficients of correlation ($\rho$) and $P$-values, are also shown in the
 Table. }
 \label{table:Table2}
  \begin{tabular}{|c|c|c|c|c|c|}

\hline

\ion{Mg}{2} vs. ${\mathrm H\beta}$ & number of AGN & $a$ & $b$ &$\rho$&P\\
\hline

${\Delta z}_{50}$ \ion{Mg}{2} vs. ${\Delta z}_{50}$ ${\mathrm H\beta}$ & 123 &  0.49$\pm$0.08 & 122.44$\pm$22.08 & 0.42 & P$<$0.00001\\
${\Delta z}_{10}$ \ion{Mg}{2} vs. ${\Delta z}_{10}$ ${\mathrm H\beta}$ & 123 & 0.10$\pm$0.08 & 791.65$\pm$120.10 & 0.10 & P$=$0.22 \\
${\Delta z}_{5}$ \ion{Mg}{2} vs. ${\Delta z}_{5}$ ${\mathrm H\beta}$ & 123 & 0.07$\pm$0.08 & 936.05$\pm$134.50 & 0.10 & P$=$0.38 \\
FWHM \ion{Mg}{2} vs. FWHM ${\mathrm H\beta}$ & 123 & 0.80$\pm$0.08 & 1425.56$\pm$336.50 & 0.68 & P$<$0.00001 \\
FW10\%M \ion{Mg}{2} vs. FW10\%M ${\mathrm H\beta}$ & 123 & 0.02$\pm$0.06 & 10530.20$\pm$919.00 & 0.04 & P$=$ 0.78 \\
FW5\%M \ion{Mg}{2} vs. FW5\%M ${\mathrm H\beta}$ & 123 & -0.05$\pm$0.06 & 13878.90$\pm$994.30 & -0.06 & P$=$ 0.36\\
\hline

\end{tabular}

\end{table*}

\appendix

\section{Appendix}

\begin{longtable}{| c | c | c | c | c | c | c |}

\caption[]{Widths and shifts of the ${\mathrm H\beta}$ line at half, 10\% and 5\% of the maximal intensity.
   \label{table:Table3}}\\

\hline  
 SDSS ID & FWHM ${\mathrm H\beta}$ & FW10\%M ${\mathrm H\beta}$ & FW5\%M ${\mathrm H\beta}$ & ${\Delta z}_{50}$ ${\mathrm H\beta}$ & ${\Delta z}_{10}$ ${\mathrm H\beta}$ & ${\Delta z}_{5}$ ${\mathrm H\beta}$  \\
(MJD-plate-fiber) & [km/s] & [km/s] & [km/s] & [km/s] & [km/s] & [km/s]\\
 
\hline
\endfirsthead
\caption{Table continued}\\
\hline 
 SDSS ID & FWHM ${\mathrm H\beta}$ & FW10\%M ${\mathrm H\beta}$ & FW5\%M ${\mathrm H\beta}$ & ${\Delta z}_{50}$ ${\mathrm H\beta}$ & ${\Delta z}_{10}$ ${\mathrm H\beta}$ & ${\Delta z}_{5}$ ${\mathrm H\beta}$ \\ 
(MJD-plate-fiber)   & [km/s] & [km/s] & [km/s] & [km/s] & [km/s] & [km/s]\\

  \hline
\endhead
\hline
\endfoot
\hline
\endlastfoot

 51633-0268-235    &    4529  &  8512  &  9941  &  -47  &  -126  &  -203  \\ 
 51699-0349-613    &    3747  &  9488  &  11890  &  69  &  634  &  678  \\ 
 51703-0353-546    &    6529  &  20272  &  25188  &  25  &  224  &  224  \\ 
 51793-0402-479    &    4009  &  8735  &  10884  &  49  &  630  &  999  \\ 
 51817-0411-381    &    3496  &  9338  &  11409  &  -19  &  -276  &  -277  \\ 
 51817-0411-519    &    3197  &  8320  &  9966  &  21  &  124  &  124  \\ 
 51821-0408-611    &    2814  &  11601  &  14022  &  83  &  522  &  522  \\ 
 51821-0423-295    &    6557  &  12004  &  13622  &  701  &  1409  &  1557  \\ 
 51871-0409-213    &    2211  &  6113  &  7429  &  16  &  -12  &  -12  \\ 
 51873-0433-422    &    4614  &  9487  &  11127  &  108  &  367  &  394  \\ 
 51877-0385-061    &    3778  &  8932  &  10547  &  107  &  306  &  306  \\ 
 51883-0436-016    &    6903  &  14191  &  16293  &  660  &  1958  &  2192  \\ 
 51908-0464-576    &    7443  &  14499  &  16693  &  274  &  1078  &  1330  \\ 
 51910-0461-074    &    4153  &  9953  &  12288  &  71  &  803  &  926  \\ 
 51910-0465-603    &    2381  &  12448  &  14899  &  108  &  401  &  401  \\ 
 51913-0469-238    &    4926  &  12420  &  15046  &  159  &  774  &  786  \\ 
 51915-0453-212    &    6036  &  14233  &  16699  &  392  &  2075  &  2351  \\ 
 51929-0413-598    &    4248  &  12787  &  16809  &  36  &  719  &  729  \\ 
 51929-0490-126    &    5408  &  12397  &  14918  &  185  &  863  &  925  \\ 
 51930-0489-402    &    4862  &  10671  &  13095  &  2  &  -16  &  -20  \\ 
 51984-0498-104    &    3364  &  8791  &  11265  &  3  &  66  &  69  \\ 
 51989-0513-340    &    3251  &  7561  &  9416  &  52  &  668  &  901  \\ 
 51993-0551-179    &    5018  &  9569  &  11118  &  74  &  305  &  413  \\ 
 52000-0554-553    &    5669  &  12244  &  14371  &  316  &  1260  &  1416  \\ 
 52017-0366-584    &    4420  &  9737  &  11771  &  93  &  768  &  987  \\ 
 52026-0593-111    &    3747  &  8542  &  10322  &  67  &  386  &  414  \\ 
 52026-0593-170    &    4216  &  9645  &  11506  &  174  &  593  &  611  \\ 
 52045-0594-073    &    4046  &  8355  &  9906  &  32  &  149  &  170  \\ 
 52056-0329-577    &    4341  &  11489  &  13865  &  133  &  527  &  529  \\ 
 52056-0603-415    &    5763  &  15082  &  17206  &  400  &  3032  &  3445  \\ 
 52059-0597-337    &    3033  &  7988  &  9911  &  36  &  490  &  504  \\ 
 52059-0597-520    &    3206  &  8227  &  9887  &  27  &  174  &  175  \\ 
 52083-0628-461    &    3705  &  10004  &  12370  &  67  &  821  &  842  \\ 
 52143-0650-126    &    6655  &  12274  &  14009  &  219  &  578  &  682  \\ 
 52145-0625-288    &    6671  &  16539  &  19657  &  1045  &  2785  &  2800  \\ 
 52145-0653-185    &    2539  &  7354  &  9029  &  7  &  -33  &  -33  \\ 
 52148-0656-282    &    2960  &  9185  &  11733  &  47  &  1103  &  1129  \\ 
 52149-0666-024    &    1974  &  6471  &  8849  &  13  &  -61  &  -62  \\ 
 52149-0666-496    &    1704  &  7763  &  9692  &  11  &  -103  &  -103  \\ 
 52164-0634-430    &    6930  &  12407  &  14102  &  454  &  863  &  951  \\ 
 52173-0644-413    &    4297  &  11285  &  13751  &  -8  &  15  &  15  \\ 
 52174-0637-259    &    4689  &  9815  &  12162  &  27  &  312  &  497  \\ 
 52174-0664-455    &    3418  &  9476  &  11323  &  42  &  124  &  124  \\ 
 52178-0640-513    &    3552  &  7741  &  9156  &  125  &  865  &  1148  \\ 
 52201-0723-500    &    3620  &  8930  &  11169  &  -16  &  -169  &  -179  \\ 
 52224-0564-368    &    3754  &  10758  &  13049  &  97  &  532  &  533  \\ 
 52224-0564-471    &    3632  &  8346  &  11170  &  28  &  304  &  456  \\ 
 52224-0722-281    &    3398  &  8883  &  11061  &  39  &  464  &  477  \\ 
 52228-0721-454    &    2810  &  6056  &  7464  &  0  &  -38  &  -50  \\ 
 52233-0753-455    &    5617  &  11426  &  13278  &  134  &  292  &  297  \\ 
 52235-0755-225    &    4926  &  11468  &  13369  &  168  &  219  &  219  \\ 
 52238-0764-053    &    5121  &  12086  &  14408  &  304  &  1372  &  1461  \\ 
 52251-0752-020    &    5706  &  10397  &  11856  &  14  &  47  &  60  \\ 
 52252-0567-558    &    4510  &  9429  &  11368  &  69  &  345  &  438  \\ 
 52252-0767-300    &    3623  &  10938  &  13895  &  112  &  1732  &  1891  \\ 
 52252-0767-453    &    4497  &  9551  &  11272  &  142  &  833  &  1036  \\ 
 52261-0741-116    &    4324  &  9780  &  12292  &  -42  &  -201  &  -241  \\ 
 52266-0555-074    &    4295  &  9721  &  11626  &  181  &  696  &  739  \\ 
 52281-0768-085    &    1889  &  6867  &  8659  &  22  &  127  &  127  \\ 
 52295-0561-618    &    6590  &  13538  &  15956  &  191  &  898  &  1111  \\ 
 52312-0526-378    &    5871  &  10356  &  11665  &  354  &  684  &  758  \\ 
 52312-0827-621    &    3371  &  7995  &  9564  &  -128  &  -415  &  -421  \\ 
 52317-0600-490    &    3761  &  8231  &  9645  &  40  &  90  &  91  \\ 
 52319-0860-583    &    3639  &  7269  &  8421  &  468  &  983  &  1036  \\ 
 52339-0856-050    &    2580  &  8013  &  10530  &  -40  &  -167  &  -167  \\ 
 52353-0507-161    &    2854  &  11612  &  14654  &  72  &  1567  &  1567  \\ 
 52353-0792-116    &    7217  &  16992  &  19832  &  -191  &  -280  &  -280  \\ 
 52368-0607-581    &    3878  &  11275  &  14100  &  15  &  60  &  60  \\ 
 52370-0596-483    &    5453  &  11308  &  13802  &  38  &  136  &  182  \\ 
 52370-0793-549    &    5802  &  13914  &  17079  &  141  &  902  &  972  \\ 
 52370-0882-376    &    3879  &  10215  &  13047  &  41  &  343  &  356  \\ 
 52376-0773-629    &    2423  &  7856  &  10447  &  6  &  -355  &  -357  \\ 
 52376-0836-262    &    3109  &  7492  &  10063  &  -36  &  -450  &  -620  \\ 
 52378-0795-381    &    5178  &  10396  &  12103  &  217  &  777  &  927  \\ 
 52378-0795-528    &    3377  &  9448  &  11829  &  9  &  215  &  217  \\ 
 52378-0892-395    &    3801  &  8805  &  11930  &  32  &  16  &  8  \\ 
 52378-0916-434    &    6742  &  11616  &  13034  &  670  &  1161  &  1278  \\ 
 52381-0510-509    &    2986  &  9821  &  12448  &  47  &  488  &  488  \\ 
 52413-0976-574    &    2048  &  6217  &  8793  &  13  &  -176  &  -181  \\ 
 52442-0984-383    &    6582  &  11810  &  13269  &  578  &  1159  &  1266  \\ 
 52443-0814-467    &    6461  &  12249  &  14130  &  -20  &  -26  &  -27  \\ 
 52466-1052-370    &    2299  &  4598  &  5735  &  23  &  170  &  367  \\ 
 52516-1054-309    &    4917  &  10524  &  12220  &  111  &  175  &  175  \\ 
 52521-0738-095    &    4473  &  15273  &  19349  &  113  &  3141  &  3531  \\ 
 52523-0987-157    &    2581  &  7902  &  10506  &  -48  &  -452  &  -457  \\ 
 52592-0896-063    &    4429  &  9538  &  11232  &  237  &  740  &  778  \\ 
 52605-0897-242    &    2948  &  9028  &  11600  &  22  &  411  &  413  \\ 
 52620-0899-626    &    2917  &  8665  &  11764  &  25  &  -50  &  -52  \\ 
 52620-0932-110    &    5442  &  11317  &  15230  &  37  &  669  &  1810  \\ 
 52636-0511-567    &    3051  &  7527  &  9269  &  40  &  153  &  158  \\ 
 52636-0939-172    &    6371  &  12590  &  14535  &  286  &  1021  &  1210  \\ 
 52636-0967-044    &    6649  &  12368  &  14203  &  79  &  325  &  428  \\ 
 52636-0999-334    &    5140  &  10751  &  12743  &  17  &  -30  &  -36  \\ 
 52642-0837-298    &    2773  &  9040  &  11121  &  99  &  912  &  913  \\ 
 52646-1186-098    &    4478  &  11183  &  13065  &  468  &  523  &  523  \\ 
 52669-0848-418    &    4564  &  11445  &  13422  &  115  &  204  &  204  \\ 
 52672-1230-503    &    2343  &  5121  &  7676  &  -44  &  -256  &  -798  \\ 
 52674-1201-225    &    2356  &  5545  &  8399  &  9  &  65  &  114  \\ 
 52703-0942-488    &    3042  &  8909  &  11852  &  -11  &  -588  &  -606  \\ 
 52703-1005-518    &    3187  &  8577  &  10814  &  41  &  132  &  134  \\ 
 52703-1199-596    &    3328  &  9042  &  11210  &  7  &  -71  &  -71  \\ 
 52709-0941-616    &    6584  &  15207  &  17746  &  378  &  2169  &  2530  \\ 
 52709-1218-201    &    3878  &  9018  &  11240  &  -22  &  25  &  32  \\ 
 52710-0993-147    &    2144  &  7820  &  9662  &  -6  &  -391  &  -391  \\ 
 52721-0914-074    &    3556  &  8730  &  10399  &  97  &  359  &  361  \\ 
 52723-1004-280    &    4949  &  11156  &  13092  &  76  &  150  &  150  \\ 
 52724-1195-098    &    4407  &  9371  &  10950  &  639  &  1020  &  1022  \\ 
 52725-0994-086    &    3045  &  7858  &  9840  &  19  &  145  &  149  \\ 
 52725-1215-150    &    3049  &  6424  &  8448  &  -17  &  -114  &  -245  \\ 
 52731-1214-077    &    3899  &  8610  &  10421  &  19  &  57  &  62  \\ 
 52731-1214-160    &    5346  &  11130  &  13036  &  129  &  367  &  381  \\ 
 52738-1167-476    &    3799  &  10704  &  12937  &  -21  &  -240  &  -240  \\ 
 52753-1171-416    &    4873  &  9913  &  11983  &  49  &  428  &  668  \\ 
 52762-1237-497    &    3468  &  7387  &  9490  &  -18  &  76  &  156  \\ 
 52765-1224-379    &    3095  &  9399  &  11980  &  24  &  292  &  293  \\ 
 52767-1329-542    &    4453  &  10882  &  13004  &  97  &  1581  &  2087  \\ 
 52781-1223-625    &    2608  &  8053  &  10055  &  -1  &  -23  &  -24  \\ 
 52790-1351-428    &    6439  &  11669  &  13300  &  566  &  1040  &  1128  \\ 
 52793-1342-157    &    5073  &  9870  &  11340  &  350  &  387  &  387  \\ 
 52821-1371-377    &    4900  &  9964  &  11563  &  299  &  498  &  501  \\ 
 52872-1402-594    &    1968  &  7401  &  9517  &  -29  &  -324  &  -324  \\ 
 52974-1271-462    &    3226  &  8777  &  10531  &  -146  &  -657  &  -658  \\ 
 52990-1429-584    &    4877  &  13393  &  15851  &  479  &  1024  &  1024  \\ 
 52990-1592-018    &    6449  &  12038  &  13730  &  274  &  742  &  873  \\ 
 52992-1431-572    &    4412  &  10629  &  13126  &  -103  &  -1150  &  -1468  \\ 
 52993-1273-348    &    5952  &  12179  &  14149  &  926  &  1995  &  2089  \\ 
 52996-1427-606    &    3707  &  9349  &  11192  &  198  &  855  &  864  \\ 
 52998-1428-558    &    3537  &  9744  &  12108  &  83  &  845  &  865  \\ 
 53003-1432-091    &    4827  &  11130  &  13577  &  -98  &  -734  &  -830  \\ 
 53033-1310-559    &    4441  &  8705  &  10279  &  36  &  137  &  185  \\ 
 53033-1598-291    &    4171  &  9163  &  11367  &  20  &  -34  &  -49  \\ 
 53033-1598-604    &    3280  &  6864  &  8687  &  27  &  150  &  260  \\ 
 53035-1433-148    &    2706  &  7530  &  10569  &  6  &  -250  &  -270  \\ 
 53035-1735-322    &    3665  &  7557  &  9190  &  44  &  399  &  627  \\ 
 53050-1362-119    &    3418  &  9073  &  11251  &  58  &  261  &  264  \\ 
 53055-1606-166    &    2974  &  8139  &  10405  &  -1  &  -319  &  -327  \\ 
 53055-1744-387    &    4022  &  8769  &  10349  &  35  &  148  &  152  \\ 
 53061-1364-185    &    3593  &  8754  &  10174  &  142  &  1388  &  1666  \\ 
 53061-1378-413    &    3083  &  8435  &  10468  &  43  &  1460  &  2047  \\ 
 53061-1378-488    &    2968  &  7957  &  10520  &  -16  &  41  &  44  \\ 
 53062-1372-488    &    4403  &  10958  &  13859  &  60  &  1457  &  2200  \\ 
 53078-1604-281    &    3594  &  7707  &  9975  &  14  &  96  &  165  \\ 
 53080-1446-135    &    6623  &  15943  &  18557  &  420  &  2574  &  2969  \\ 
 53083-1367-629    &    4913  &  11283  &  13789  &  95  &  668  &  754  \\ 
 53084-1380-058    &    6258  &  11508  &  13171  &  64  &  179  &  228  \\ 
 53084-1440-137    &    3739  &  9581  &  12040  &  27  &  523  &  554  \\ 
 53084-1453-010    &    5534  &  14434  &  17698  &  167  &  2177  &  2782  \\ 
 53084-1619-060    &    5505  &  12617  &  15061  &  140  &  562  &  578  \\ 
 53089-1455-584    &    6634  &  11923  &  13323  &  689  &  1304  &  1378  \\ 
 53089-1779-106    &    3437  &  10220  &  13141  &  26  &  648  &  655  \\ 
 53108-1394-044    &    4423  &  8836  &  10703  &  -24  &  -120  &  -195  \\ 
 53112-1773-301    &    4652  &  11691  &  13766  &  -70  &  -141  &  -141  \\ 
 53112-1773-405    &    2320  &  7144  &  10141  &  -5  &  299  &  306  \\ 
 53116-1457-421    &    3863  &  8915  &  11243  &  -57  &  -432  &  -528  \\ 
 53119-1603-558    &    5499  &  10275  &  11889  &  1009  &  1436  &  1444  \\ 
 53141-1417-441    &    5363  &  10654  &  12432  &  170  &  610  &  733  \\ 
 53147-1676-055    &    3479  &  9239  &  12704  &  13  &  619  &  725  \\ 
 53172-1645-526    &    5824  &  11780  &  13690  &  158  &  323  &  330  \\ 
 53228-1728-218    &    2805  &  6838  &  9219  &  -13  &  205  &  266  \\ 
 53312-1865-572    &    3440  &  9429  &  12900  &  44  &  1576  &  2777  \\ 
 53313-1864-353    &    4467  &  10215  &  12264  &  20  &  133  &  138  \\ 
 53317-1926-378    &    2749  &  7140  &  9574  &  -23  &  -467  &  -542  \\ 
 53350-2080-309    &    3709  &  8866  &  11182  &  -16  &  -262  &  -295  \\ 
 53358-1750-564    &    4986  &  10935  &  12981  &  163  &  927  &  1066  \\ 
 53383-1940-407    &    7203  &  14543  &  16783  &  404  &  502  &  502  \\ 
 53385-1754-324    &    3171  &  9330  &  12310  &  39  &  187  &  188  \\ 
 53385-1944-412    &    3290  &  9093  &  11017  &  70  &  265  &  265  \\ 
 53386-1872-371    &    3946  &  11269  &  13944  &  163  &  1688  &  1810  \\ 
 53386-1941-545    &    3730  &  9149  &  11905  &  52  &  508  &  603  \\ 
 53430-1667-114    &    3736  &  9470  &  11546  &  -76  &  -573  &  -586  \\ 
 53431-2020-614    &    6260  &  13653  &  15805  &  439  &  439  &  439  \\ 
 53433-1949-437    &    2983  &  8754  &  10518  &  13  &  -14  &  -14  \\ 
 53433-1949-472    &    1855  &  9323  &  13225  &  37  &  1847  &  1850  \\ 
 53436-1683-016    &    4877  &  15006  &  17883  &  106  &  3179  &  3897  \\ 
 53437-0614-452    &    3132  &  7694  &  10396  &  4  &  444  &  596  \\ 
 53442-2003-501    &    3267  &  7539  &  9672  &  18  &  352  &  469  \\ 
 53466-2034-230    &    3729  &  8693  &  10578  &  28  &  220  &  233  \\ 
 53467-1838-451    &    6517  &  13961  &  16390  &  -117  &  -358  &  -365  \\ 
 53472-1626-449    &    3992  &  12120  &  14942  &  104  &  2259  &  2653  \\ 
 53472-1694-540    &    3108  &  6500  &  8659  &  -26  &  -75  &  -160  \\ 
 53473-1695-612    &    2310  &  7193  &  9909  &  -47  &  -575  &  -588  \\ 
 53473-2008-615    &    3332  &  8648  &  10544  &  106  &  601  &  610  \\ 
 53474-2007-171    &    5446  &  14432  &  16359  &  308  &  2885  &  3250  \\ 
 53474-2007-226    &    3296  &  11160  &  13845  &  -44  &  -415  &  -415  \\ 
 53475-1690-634    &    6060  &  13950  &  16990  &  181  &  1167  &  1369  \\ 
 53493-1575-484    &    2039  &  6919  &  8481  &  -58  &  -300  &  -300  \\ 
 53498-2089-385    &    9030  &  16304  &  18540  &  566  &  1137  &  1265  \\ 
 53499-2030-201    &    5754  &  11939  &  13971  &  168  &  1067  &  1375  \\ 
 53521-1714-011    &    3642  &  13992  &  18377  &  77  &  1967  &  1972  \\ 
 53521-1714-571    &    3783  &  9825  &  12029  &  103  &  637  &  649  \\ 
 53566-1854-151    &    2916  &  8035  &  11041  &  4  &  909  &  1170  \\ 
 53674-2265-087    &    3312  &  8936  &  11112  &  42  &  533  &  544  \\ 
 53711-2281-431    &    5194  &  10037  &  11582  &  168  &  598  &  715  \\ 
 53712-2276-508    &    3472  &  8168  &  12484  &  17  &  649  &  1765  \\ 
 53713-2292-348    &    6651  &  14092  &  16454  &  530  &  1012  &  1017  \\ 
 53714-2106-130    &    4083  &  8154  &  9722  &  62  &  431  &  661  \\ 
 53733-2294-607    &    4488  &  11389  &  13440  &  -444  &  -913  &  -913  \\ 
 53737-2364-469    &    2714  &  6131  &  8139  &  23  &  9  &  1  \\ 
 53738-2341-066    &    3235  &  8168  &  10453  &  42  &  241  &  257  \\ 
 53739-2365-591    &    2813  &  8372  &  10868  &  -4  &  -442  &  -447  \\ 
 53757-2347-151    &    3522  &  12496  &  15725  &  78  &  2782  &  3226  \\ 
 53757-2347-326    &    4386  &  9485  &  11343  &  66  &  381  &  425  \\ 
 53759-2362-295    &    3372  &  8477  &  10805  &  -3  &  295  &  322  \\ 
 53770-2352-242    &    3645  &  9633  &  12624  &  72  &  751  &  832  \\ 
 53770-2376-497    &    2399  &  6528  &  8630  &  10  &  40  &  41  \\ 
 53772-1989-559    &    1583  &  6753  &  8811  &  -30  &  -198  &  -198  \\ 
 53772-2351-099    &    4924  &  11437  &  13787  &  -62  &  -376  &  -392  \\ 
 53772-2351-616    &    4035  &  8335  &  9995  &  6  &  137  &  177  \\ 
 53794-1810-504    &    3080  &  8472  &  11020  &  16  &  -99  &  -102  \\ 
 53794-2353-600    &    4276  &  9578  &  11524  &  117  &  665  &  748  \\ 
 53795-2216-298    &    4100  &  10627  &  12740  &  273  &  1586  &  1685  \\ 
 53799-2222-388    &    4570  &  11474  &  13782  &  114  &  1729  &  2247  \\ 
 53800-2128-415    &    3620  &  8258  &  10658  &  -6  &  -11  &  -12  \\ 
 53801-2428-356    &    6023  &  15080  &  18181  &  295  &  1942  &  2130  \\ 
 53816-1843-008    &    3104  &  8339  &  10004  &  -60  &  -254  &  -255  \\ 
 53816-1843-502    &    4866  &  10313  &  12797  &  -70  &  -426  &  -634  \\ 
 53816-1843-584    &    4100  &  9768  &  12374  &  66  &  855  &  1139  \\ 
 53818-2440-539    &    5538  &  15199  &  18502  &  265  &  2316  &  2504  \\ 
 53827-1713-063    &    4729  &  12519  &  15393  &  118  &  1050  &  1078  \\ 
 53827-1716-350    &    5427  &  13849  &  16481  &  -136  &  -372  &  -372  \\ 
 53828-2237-093    &    4439  &  8636  &  9964  &  308  &  456  &  458  \\ 
 53845-2233-196    &    5268  &  12601  &  14816  &  -7  &  -24  &  -24  \\ 
 53847-1775-420    &    2858  &  8752  &  12531  &  -30  &  -1052  &  -1170  \\ 
 53851-1817-069    &    3059  &  5585  &  6378  &  11  &  11  &  11  \\ 
 53851-2023-065    &    3306  &  10428  &  13413  &  -5  &  -92  &  -92  \\ 
 53855-2491-252    &    5542  &  14495  &  16840  &  303  &  2681  &  3108  \\ 
 53858-2101-348    &    2532  &  7408  &  9796  &  22  &  -129  &  -131  \\ 
 53874-2152-108    &    4647  &  10771  &  13112  &  124  &  821  &  928  \\ 
 53874-2173-354    &    4232  &  9454  &  11657  &  90  &  534  &  658  \\ 
 53875-2508-353    &    4565  &  8328  &  9464  &  104  &  304  &  355  \\ 
 53877-2512-582    &    5554  &  11821  &  14166  &  121  &  865  &  1144  \\ 
 53883-1793-573    &    3354  &  11508  &  15301  &  87  &  2133  &  2321  \\ 
 53884-1796-591    &    6373  &  12584  &  14563  &  685  &  1487  &  1570  \\ 
 53886-2164-503    &    3605  &  9520  &  11512  &  65  &  455  &  457  \\ 
 54084-2501-120    &    4594  &  11143  &  13378  &  193  &  900  &  923  \\ 
 54086-2420-240    &    3985  &  10047  &  12181  &  21  &  175  &  177  \\ 
 54095-2583-462    &    3675  &  8779  &  10676  &  22  &  96  &  99  \\ 
 54095-2583-615    &    2966  &  9194  &  11611  &  64  &  434  &  434  \\ 
 54097-2478-377    &    2376  &  7985  &  10703  &  5  &  -337  &  -338  \\ 
 54115-2102-018    &    4020  &  9604  &  11320  &  123  &  299  &  299  \\ 
 54115-2493-278    &    3952  &  8670  &  11756  &  6  &  270  &  567  \\ 
 54138-1844-112    &    5449  &  10751  &  12477  &  172  &  770  &  991  \\ 
 54139-2425-442    &    5851  &  14929  &  17564  &  559  &  2712  &  2910  \\ 
 54144-1845-637    &    6106  &  14792  &  17466  &  128  &  328  &  328  \\ 
 54149-2488-521    &    4555  &  14749  &  18427  &  97  &  1167  &  1168  \\ 
 54152-1803-323    &    5269  &  11049  &  12612  &  387  &  1483  &  1708  \\ 
 54153-2242-199    &    4894  &  10187  &  12351  &  -45  &  -205  &  -266  \\ 
 54154-2606-614    &    3426  &  9002  &  11900  &  -36  &  -573  &  -637  \\ 
 54175-1807-009    &    2316  &  6996  &  8787  &  -10  &  -37  &  -37  \\ 
 54176-1847-630    &    4231  &  8969  &  10912  &  32  &  221  &  282  \\ 
 54176-2485-231    &    10334  &  18710  &  21287  &  411  &  937  &  1066  \\ 
 54178-2492-562    &    5270  &  9605  &  10957  &  2  &  2  &  3  \\ 
 54178-2592-153    &    3410  &  8351  &  10949  &  -1  &  342  &  414  \\ 
 54180-2509-244    &    3618  &  8947  &  11372  &  39  &  1007  &  1429  \\ 
 54207-2595-541    &    5342  &  9998  &  11435  &  182  &  556  &  668  \\ 
 54230-2641-570    &    5124  &  11213  &  13331  &  185  &  825  &  911  \\ 
 54232-2166-030    &    2423  &  7033  &  9706  &  -46  &  -455  &  -475  \\ 
 54243-2518-374    &    3694  &  11319  &  13489  &  241  &  2369  &  2549  \\ 
 54266-1725-540    &    2861  &  9653  &  11847  &  86  &  539  &  539  \\ 
 54266-1791-046    &    2527  &  8216  &  11470  &  13  &  166  &  167  \\ 
 54272-2744-535    &    2554  &  7816  &  11029  &  -9  &  -616  &  -643  \\ 
 54448-2424-109    &    2019  &  8864  &  11050  &  38  &  484  &  484  \\ 
 54477-2645-133    &    3788  &  9030  &  11006  &  80  &  349  &  364  \\ 
 54479-2646-204    &    4692  &  10819  &  12705  &  225  &  1431  &  1688  \\ 
 54481-2613-620    &    2492  &  6170  &  9247  &  -26  &  24  &  47  \\ 
 54483-2615-257    &    2218  &  6423  &  8375  &  24  &  -130  &  -132  \\ 
 54483-2615-404    &    3303  &  11501  &  14611  &  109  &  1619  &  1629  \\ 
 54495-2647-160    &    5341  &  11647  &  14659  &  10  &  214  &  303  \\ 
 54497-2885-561    &    4663  &  8732  &  9981  &  142  &  464  &  562  \\ 
 54502-2617-287    &    2819  &  7185  &  8829  &  78  &  430  &  441  \\ 
 54507-1795-509    &    3777  &  12110  &  13977  &  107  &  2816  &  3287  \\ 
 54523-2151-527    &    2305  &  7501  &  9879  &  19  &  586  &  590  \\ 
 54524-2783-432    &    4980  &  9280  &  10507  &  353  &  888  &  1006  \\ 
 54527-2769-354    &    2948  &  9305  &  12477  &  -17  &  -292  &  -294  \\ 
 54529-2772-038    &    3151  &  8809  &  11521  &  16  &  804  &  867  \\ 
 54534-2774-051    &    7294  &  14911  &  17025  &  727  &  2158  &  2431  \\ 
 54535-2764-053    &    3093  &  8250  &  10684  &  45  &  713  &  785  \\ 
 54562-2955-608    &    3447  &  8698  &  11256  &  -23  &  -37  &  -38  \\ 
 54563-2795-140    &    6120  &  12639  &  14694  &  655  &  1001  &  1003  \\ 
 54567-2517-624    &    2293  &  8041  &  9969  &  2  &  -237  &  -237  \\ 
 54570-2522-131    &    4064  &  10894  &  14375  &  76  &  1781  &  2911  \\ 
 54582-2526-404    &    3706  &  8232  &  10260  &  34  &  597  &  895  \\ 
 54584-2520-442    &    6040  &  13853  &  16167  &  841  &  1095  &  1095  \\ 
 54585-2529-300    &    3514  &  9254  &  11842  &  17  &  315  &  328  \\ 
 54589-2532-090    &    4530  &  10194  &  12457  &  92  &  776  &  978  \\ 
 54589-2970-373    &    3194  &  10759  &  13867  &  64  &  993  &  995  \\ 
 54590-2971-326    &    2342  &  7923  &  10079  &  39  &  242  &  242  \\   \hline

\end{longtable}

\pagebreak

\begin{longtable}{| c | c | c | c | c | c | c |}
\caption[]{Widths and shifts of the \ion{Mg}{2} line at half, 10\% and 5\% of the maximal intensity.
   \label{table:Table4}}\\

\hline 
 SDSS ID & FWHM \ion{Mg}{2} & FW10\%M \ion{Mg}{2} & FW5\%M \ion{Mg}{2} & ${\Delta z}_{50}$ \ion{Mg}{2} & ${\Delta z}_{10}$ \ion{Mg}{2} & ${\Delta z}_{5}$ \ion{Mg}{2}  \\
 (MJD-plate-fiber) & [km/s] & [km/s] & [km/s] & [km/s] & [km/s] & [km/s]\\

\hline
\endfirsthead
\caption{Table continued}\\
\hline
 SDSS ID & FWHM \ion{Mg}{2} & FW10\%M \ion{Mg}{2} & FW5\%M \ion{Mg}{2} & ${\Delta z}_{50}$ \ion{Mg}{2} & ${\Delta z}_{10}$ \ion{Mg}{2} & ${\Delta z}_{5}$ \ion{Mg}{2} \\ 
(MJD-plate-fiber)  & [km/s] & [km/s] & [km/s] & [km/s] & [km/s] & [km/s]\\
 \hline
\endhead
\hline
\endfoot
\hline
\endlastfoot
 51633-0268-235    &    4951  &  17580  &  24989  &  44  &  297  &  297  \\ 
 51699-0349-613    &    2875  &  24042  &  24579  &  40  &  605  &  605  \\ 
 51703-0353-546    &    6566  &  21014  &  24335  &  -79  &  -337  &  -336  \\ 
 51793-0402-479    &    3617  &  14447  &  23860  &  -58  &  -762  &  -762  \\ 
 51817-0411-381    &    3059  &  14447  &  25233  &  25  &  545  &  545  \\ 
 51817-0411-519    &    3139  &  21863  &  21923  &  -3  &  1157  &  1157  \\ 
 51821-0408-611    &    3123  &  21874  &  21310  &  -1  &  -947  &  -947  \\ 
 51821-0423-295    &    6051  &  20659  &  15409  &  1410  &  2150  &  2150  \\ 
 51871-0409-213    &    2120  &  20418  &  19894  &  -15  &  368  &  368  \\ 
 51873-0433-422    &    4085  &  18678  &  14447  &  226  &  2198  &  2379  \\ 
 51877-0385-061    &    3450  &  26471  &  21818  &  37  &  260  &  260  \\ 
 51883-0436-016    &    4409  &  23526  &  16296  &  240  &  2853  &  3140  \\ 
 51908-0464-576    &    4639  &  26920  &  16555  &  345  &  2030  &  2047  \\ 
 51910-0461-074    &    4377  &  14606  &  25667  &  60  &  773  &  774  \\ 
 51910-0465-603    &    4085  &  21535  &  14447  &  226  &  2198  &  2379  \\ 
 51913-0469-238    &    5032  &  25193  &  22310  &  47  &  235  &  235  \\ 
 51915-0453-212    &    6747  &  21299  &  17121  &  419  &  1861  &  2140  \\ 
 51929-0413-598    &    3368  &  14447  &  12869  &  104  &  2413  &  2793  \\ 
 51929-0490-126    &    4095  &  14447  &  14364  &  182  &  2457  &  2827  \\ 
 51930-0489-402    &    4596  &  16612  &  28439  &  -51  &  -679  &  -679  \\ 
 51984-0498-104    &    3016  &  25830  &  14192  &  25  &  339  &  340  \\ 
 51989-0513-340    &    4085  &  24166  &  14447  &  226  &  2198  &  2379  \\ 
 51993-0551-179    &    4091  &  19083  &  23680  &  23  &  212  &  212  \\ 
 52000-0554-553    &    5011  &  26218  &  22152  &  6  &  270  &  270  \\ 
 52017-0366-584    &    4085  &  24263  &  14447  &  226  &  2198  &  2379  \\ 
 52026-0593-111    &    3279  &  14155  &  12749  &  70  &  1410  &  1536  \\ 
 52026-0593-170    &    3067  &  14517  &  14367  &  64  &  2696  &  2941  \\ 
 52045-0594-073    &    3322  &  19302  &  18898  &  -31  &  -241  &  -241  \\ 
 52056-0329-577    &    4085  &  23311  &  14447  &  226  &  2198  &  2379  \\ 
 52056-0603-415    &    5869  &  20308  &  14948  &  294  &  1803  &  2079  \\ 
 52059-0597-337    &    2894  &  21500  &  23123  &  66  &  1763  &  1763  \\ 
 52059-0597-520    &    2422  &  13115  &  19016  &  6  &  15  &  16  \\ 
 52083-0628-461    &    2892  &  15223  &  21336  &  11  &  194  &  194  \\ 
 52143-0650-126    &    5809  &  18844  &  21391  &  -8  &  -102  &  -103  \\ 
 52145-0625-288    &    4649  &  24047  &  17240  &  277  &  2550  &  2654  \\ 
 52145-0653-185    &    1831  &  25772  &  11428  &  -3  &  691  &  926  \\ 
 52148-0656-282    &    2337  &  22659  &  10803  &  62  &  2109  &  2623  \\ 
 52149-0666-024    &    2307  &  21106  &  24398  &  11  &  836  &  836  \\ 
 52149-0666-496    &    2204  &  20249  &  22611  &  -14  &  -338  &  -338  \\ 
 52164-0634-430    &    4604  &  24563  &  13731  &  333  &  2414  &  2744  \\ 
 52173-0644-413    &    4085  &  24247  &  23098  &  -80  &  -1527  &  -1526  \\ 
 52174-0637-259    &    3382  &  23226  &  24350  &  24  &  -240  &  -240  \\ 
 52174-0664-455    &    3196  &  20882  &  26189  &  8  &  47  &  47  \\ 
 52178-0640-513    &    3266  &  24635  &  14608  &  128  &  2390  &  2517  \\ 
 52201-0723-500    &    2929  &  24095  &  20385  &  6  &  -776  &  -775  \\ 
 52224-0564-368    &    4085  &  19846  &  14447  &  226  &  2198  &  2379  \\ 
 52224-0564-471    &    4085  &  18836  &  14447  &  226  &  2198  &  2379  \\ 
 52224-0722-281    &    3704  &  21399  &  19450  &  -11  &  -308  &  -308  \\ 
 52228-0721-454    &    2819  &  21596  &  25271  &  23  &  -524  &  -524  \\ 
 52233-0753-455    &    4085  &  18816  &  14447  &  226  &  2198  &  2379  \\ 
 52235-0755-225    &    4085  &  24636  &  14447  &  226  &  2198  &  2379  \\ 
 52238-0764-053    &    4621  &  22038  &  21259  &  55  &  453  &  454  \\ 
 52251-0752-020    &    4395  &  21767  &  26138  &  14  &  -91  &  -91  \\ 
 52252-0567-558    &    3597  &  21087  &  23643  &  38  &  520  &  521  \\ 
 52252-0767-300    &    4933  &  14383  &  23536  &  72  &  188  &  188  \\ 
 52252-0767-453    &    3862  &  12732  &  23907  &  -21  &  -697  &  -697  \\ 
 52261-0741-116    &    4113  &  26062  &  22213  &  -93  &  -1383  &  -1383  \\ 
 52266-0555-074    &    3894  &  26002  &  13288  &  198  &  1892  &  2057  \\ 
 52281-0768-085    &    2193  &  22518  &  19718  &  -1  &  468  &  469  \\ 
 52295-0561-618    &    5927  &  18057  &  13415  &  472  &  1117  &  1212  \\ 
 52312-0526-378    &    3766  &  14447  &  21367  &  -12  &  -243  &  -242  \\ 
 52312-0827-621    &    2031  &  21922  &  21849  &  -29  &  -574  &  -575  \\ 
 52317-0600-490    &    2964  &  22361  &  14306  &  23  &  1510  &  1523  \\ 
 52319-0860-583    &    2980  &  20824  &  26633  &  -16  &  -83  &  -83  \\ 
 52339-0856-050    &    2348  &  16346  &  23580  &  -39  &  -1504  &  -1504  \\ 
 52353-0507-161    &    3077  &  17594  &  19279  &  -23  &  6  &  6  \\ 
 52353-0792-116    &    2964  &  20911  &  20001  &  -26  &  -244  &  -244  \\ 
 52368-0607-581    &    4090  &  19162  &  21729  &  -20  &  73  &  72  \\ 
 52370-0596-483    &    4781  &  20372  &  21926  &  -33  &  -156  &  -156  \\ 
 52370-0793-549    &    4398  &  19409  &  22810  &  10  &  69  &  69  \\ 
 52370-0882-376    &    2444  &  17308  &  20347  &  -29  &  -679  &  -679  \\ 
 52376-0773-629    &    2070  &  19174  &  17857  &  7  &  -882  &  -883  \\ 
 52376-0836-262    &    2711  &  21614  &  20254  &  7  &  265  &  265  \\ 
 52378-0795-381    &    4681  &  14495  &  25020  &  2  &  354  &  354  \\ 
 52378-0795-528    &    3368  &  22527  &  21709  &  -52  &  -794  &  -794  \\ 
 52378-0892-395    &    4046  &  21747  &  19897  &  46  &  505  &  506  \\ 
 52378-0916-434    &    5082  &  25260  &  15374  &  309  &  1736  &  1820  \\ 
 52381-0510-509    &    2945  &  23911  &  18564  &  -13  &  -941  &  -941  \\ 
 52413-0976-574    &    1862  &  21586  &  16958  &  -24  &  677  &  677  \\ 
 52442-0984-383    &    5561  &  20362  &  24043  &  70  &  321  &  321  \\ 
 52443-0814-467    &    3711  &  10624  &  22278  &  -25  &  -388  &  -388  \\ 
 52466-1052-370    &    4085  &  19336  &  14447  &  226  &  2198  &  2379  \\ 
 52516-1054-309    &    4125  &  16115  &  26033  &  -68  &  -444  &  -444  \\ 
 52521-0738-095    &    4085  &  24059  &  14447  &  226  &  2198  &  2379  \\ 
 52523-0987-157    &    2472  &  9864  &  21124  &  -26  &  -1749  &  -1749  \\ 
 52592-0896-063    &    4219  &  19010  &  23391  &  6  &  273  &  273  \\ 
 52605-0897-242    &    2478  &  28281  &  21585  &  3  &  215  &  215  \\ 
 52620-0899-626    &    2865  &  12879  &  26912  &  6  &  -247  &  -247  \\ 
 52620-0932-110    &    5192  &  22851  &  25566  &  0  &  -175  &  -175  \\ 
 52636-0511-567    &    2610  &  20783  &  21977  &  14  &  -852  &  -852  \\ 
 52636-0939-172    &    6681  &  26792  &  23683  &  8  &  32  &  32  \\ 
 52636-0967-044    &    3707  &  24120  &  13567  &  83  &  1546  &  1661  \\ 
 52636-0999-334    &    4927  &  20947  &  13650  &  229  &  1859  &  2214  \\ 
 52642-0837-298    &    3594  &  23403  &  15590  &  101  &  2252  &  2343  \\ 
 52646-1186-098    &    3995  &  23614  &  12530  &  153  &  1685  &  1990  \\ 
 52669-0848-418    &    4062  &  19350  &  15026  &  216  &  2138  &  2234  \\ 
 52672-1230-503    &    2542  &  24608  &  19098  &  34  &  221  &  221  \\ 
 52674-1201-225    &    2056  &  21762  &  20719  &  14  &  -146  &  -147  \\ 
 52703-0942-488    &    2430  &  20461  &  20325  &  4  &  1890  &  1891  \\ 
 52703-1005-518    &    2883  &  14515  &  21125  &  -55  &  -1172  &  -1172  \\ 
 52703-1199-596    &    2304  &  21343  &  17580  &  -30  &  495  &  496  \\ 
 52709-0941-616    &    4031  &  19856  &  24042  &  -41  &  -222  &  -222  \\ 
 52709-1218-201    &    3035  &  19231  &  21014  &  -20  &  -830  &  -830  \\ 
 52710-0993-147    &    4085  &  22895  &  14447  &  226  &  2198  &  2379  \\ 
 52721-0914-074    &    4085  &  19744  &  14447  &  226  &  2198  &  2379  \\ 
 52723-1004-280    &    3572  &  20407  &  21863  &  -61  &  -1249  &  -1249  \\ 
 52724-1195-098    &    3663  &  22393  &  21874  &  -7  &  31  &  31  \\ 
 52725-0994-086    &    2201  &  23650  &  20659  &  -4  &  282  &  281  \\ 
 52725-1215-150    &    2274  &  19637  &  20418  &  -35  &  -952  &  -951  \\ 
 52731-1214-077    &    2981  &  12216  &  18678  &  -54  &  -2771  &  -2879  \\ 
 52731-1214-160    &    3503  &  22134  &  26471  &  -19  &  14  &  14  \\ 
 52738-1167-476    &    3623  &  21070  &  23526  &  18  &  423  &  422  \\ 
 52753-1171-416    &    4574  &  13105  &  26920  &  -95  &  -835  &  -834  \\ 
 52762-1237-497    &    3468  &  21034  &  14606  &  64  &  1465  &  1503  \\ 
 52765-1224-379    &    3273  &  22946  &  21535  &  -35  &  -275  &  -274  \\ 
 52767-1329-542    &    3503  &  23493  &  25193  &  7  &  315  &  315  \\ 
 52781-1223-625    &    3078  &  15589  &  21299  &  23  &  -303  &  -303  \\ 
 52790-1351-428    &    4085  &  22599  &  14447  &  226  &  2198  &  2379  \\ 
 52793-1342-157    &    4085  &  19620  &  14447  &  226  &  2198  &  2379  \\ 
 52821-1371-377    &    3733  &  23656  &  16612  &  196  &  2366  &  2372  \\ 
 52872-1402-594    &    2100  &  20726  &  25830  &  -30  &  -884  &  -884  \\ 
 52974-1271-462    &    2528  &  23159  &  24166  &  -18  &  1227  &  1227  \\ 
 52990-1429-584    &    4391  &  25508  &  19083  &  69  &  1020  &  1021  \\ 
 52990-1592-018    &    3394  &  18676  &  26218  &  -45  &  -557  &  -558  \\ 
 52992-1431-572    &    3211  &  25084  &  24263  &  33  &  -32  &  -31  \\ 
 52993-1273-348    &    5782  &  13282  &  14155  &  998  &  2153  &  2225  \\ 
 52996-1427-606    &    4779  &  21926  &  14517  &  347  &  2715  &  3040  \\ 
 52998-1428-558    &    3339  &  25580  &  19302  &  12  &  -553  &  -553  \\ 
 53003-1432-091    &    4852  &  18929  &  23311  &  -80  &  -678  &  -679  \\ 
 53033-1310-559    &    4013  &  17747  &  20308  &  -66  &  -1435  &  -1435  \\ 
 53033-1598-291    &    4623  &  19081  &  21500  &  66  &  926  &  926  \\ 
 53033-1598-604    &    2912  &  16970  &  13115  &  43  &  759  &  759  \\ 
 53035-1433-148    &    2117  &  18800  &  15223  &  7  &  292  &  292  \\ 
 53035-1735-322    &    2752  &  12832  &  18844  &  -20  &  880  &  880  \\ 
 53050-1362-119    &    3065  &  15254  &  24047  &  -33  &  -290  &  -291  \\ 
 53055-1606-166    &    3074  &  11951  &  25772  &  -6  &  453  &  453  \\ 
 53055-1744-387    &    3766  &  21672  &  22659  &  -18  &  -665  &  -665  \\ 
 53061-1364-185    &    3320  &  11669  &  21106  &  2  &  508  &  509  \\ 
 53061-1378-413    &    2635  &  16515  &  20249  &  16  &  -626  &  -626  \\ 
 53061-1378-488    &    2891  &  21821  &  24563  &  -6  &  334  &  335  \\ 
 53062-1372-488    &    3909  &  33760  &  24247  &  2  &  -736  &  -736  \\ 
 53078-1604-281    &    2648  &  15143  &  23226  &  0  &  -661  &  -661  \\ 
 53080-1446-135    &    4649  &  15906  &  20882  &  -30  &  -273  &  -273  \\ 
 53083-1367-629    &    3890  &  19109  &  24635  &  1  &  -7  &  -6  \\ 
 53084-1380-058    &    6749  &  10569  &  24095  &  -68  &  -261  &  -261  \\ 
 53084-1440-137    &    2623  &  21326  &  19846  &  -39  &  -731  &  -731  \\ 
 53084-1453-010    &    3168  &  21937  &  18836  &  4  &  200  &  200  \\ 
 53084-1619-060    &    3808  &  20902  &  21399  &  -9  &  215  &  215  \\ 
 53089-1455-584    &    4355  &  20815  &  21596  &  6  &  137  &  137  \\ 
 53089-1779-106    &    3310  &  19409  &  18816  &  15  &  -381  &  -381  \\ 
 53108-1394-044    &    3192  &  25958  &  24636  &  -2  &  721  &  721  \\ 
 53112-1773-301    &    3122  &  24010  &  22038  &  27  &  459  &  459  \\ 
 53112-1773-405    &    2404  &  23418  &  21767  &  1  &  131  &  132  \\ 
 53116-1457-421    &    3574  &  12726  &  21087  &  -20  &  -435  &  -436  \\ 
 53119-1603-558    &    4470  &  20642  &  14383  &  434  &  2511  &  2680  \\ 
 53141-1417-441    &    4806  &  13114  &  12732  &  189  &  1413  &  1751  \\ 
 53147-1676-055    &    2767  &  21841  &  26062  &  23  &  319  &  320  \\ 
 53172-1645-526    &    6731  &  19818  &  26002  &  -25  &  -101  &  -101  \\ 
 53228-1728-218    &    2558  &  24090  &  22518  &  -28  &  379  &  379  \\ 
 53312-1865-572    &    1831  &  24033  &  18057  &  4  &  687  &  687  \\ 
 53313-1864-353    &    4085  &  22208  &  14447  &  226  &  2198  &  2379  \\ 
 53317-1926-378    &    2703  &  19486  &  21922  &  -16  &  -164  &  -164  \\ 
 53350-2080-309    &    3243  &  13309  &  22361  &  20  &  -44  &  -43  \\ 
 53358-1750-564    &    4111  &  12876  &  20824  &  60  &  395  &  395  \\ 
 53383-1940-407    &    6341  &  23989  &  16346  &  476  &  1591  &  1698  \\ 
 53385-1754-324    &    2365  &  19889  &  17594  &  21  &  877  &  877  \\ 
 53385-1944-412    &    3680  &  22540  &  20911  &  9  &  614  &  614  \\ 
 53386-1872-371    &    3852  &  15745  &  19162  &  10  &  114  &  114  \\ 
 53386-1941-545    &    2722  &  26586  &  20372  &  -3  &  -1024  &  -1023  \\ 
 53430-1667-114    &    2951  &  26141  &  19409  &  17  &  584  &  584  \\ 
 53431-2020-614    &    4586  &  21896  &  17308  &  325  &  902  &  902  \\ 
 53433-1949-437    &    2264  &  21479  &  19174  &  15  &  345  &  345  \\ 
 53433-1949-472    &    6425  &  12423  &  21614  &  1047  &  1362  &  1363  \\ 
 53436-1683-016    &    5020  &  14880  &  14495  &  228  &  1424  &  1549  \\ 
 53437-0614-452    &    2706  &  16580  &  22527  &  22  &  322  &  321  \\ 
 53442-2003-501    &    2600  &  22415  &  21747  &  -12  &  462  &  462  \\ 
 53466-2034-230    &    3416  &  25741  &  25260  &  24  &  131  &  131  \\ 
 53467-1838-451    &    3625  &  21764  &  23911  &  19  &  -7  &  -7  \\ 
 53472-1626-449    &    3940  &  22901  &  21586  &  23  &  682  &  682  \\ 
 53472-1694-540    &    2383  &  21982  &  20362  &  -1  &  714  &  714  \\ 
 53473-1695-612    &    2070  &  22099  &  10624  &  17  &  2562  &  3188  \\ 
 53473-2008-615    &    2747  &  23874  &  19336  &  16  &  136  &  135  \\ 
 53474-2007-171    &    4934  &  14496  &  16115  &  309  &  1702  &  1729  \\ 
 53474-2007-226    &    2743  &  12863  &  24059  &  -29  &  259  &  259  \\ 
 53475-1690-634    &    3482  &  21944  &  9864  &  538  &  1580  &  1615  \\ 
 53493-1575-484    &    2322  &  14476  &  19010  &  30  &  775  &  776  \\ 
 53498-2089-385    &    8503  &  19455  &  28281  &  279  &  1165  &  1166  \\ 
 53499-2030-201    &    4620  &  10898  &  12879  &  728  &  1991  &  2035  \\ 
 53521-1714-011    &    3366  &  17366  &  22851  &  10  &  -598  &  -598  \\ 
 53521-1714-571    &    3583  &  16781  &  20783  &  -5  &  -255  &  -254  \\ 
 53566-1854-151    &    3254  &  21235  &  26792  &  -7  &  -322  &  -321  \\ 
 53674-2265-087    &    3611  &  18850  &  24120  &  -25  &  -580  &  -580  \\ 
 53711-2281-431    &    4923  &  17370  &  20947  &  67  &  286  &  286  \\ 
 53712-2276-508    &    3236  &  17362  &  23403  &  -32  &  -769  &  -770  \\ 
 53713-2292-348    &    5760  &  19793  &  23614  &  -205  &  -596  &  -596  \\ 
 53714-2106-130    &    2280  &  13043  &  19350  &  -4  &  724  &  724  \\ 
 53733-2294-607    &    2217  &  17652  &  24608  &  19  &  961  &  962  \\ 
 53737-2364-469    &    2290  &  13798  &  21762  &  21  &  -103  &  -103  \\ 
 53738-2341-066    &    2835  &  15746  &  20461  &  16  &  19  &  19  \\ 
 53739-2365-591    &    2382  &  10312  &  14515  &  -13  &  1359  &  1361  \\ 
 53757-2347-151    &    2894  &  16159  &  21343  &  8  &  -216  &  -216  \\ 
 53757-2347-326    &    3197  &  15542  &  19856  &  16  &  345  &  345  \\ 
 53759-2362-295    &    1998  &  12669  &  19231  &  -16  &  1308  &  1308  \\ 
 53770-2352-242    &    3357  &  17929  &  22895  &  -22  &  128  &  128  \\ 
 53770-2376-497    &    2619  &  13949  &  19744  &  7  &  141  &  141  \\ 
 53772-1989-559    &    1956  &  13120  &  20407  &  17  &  -743  &  -744  \\ 
 53772-2351-099    &    3441  &  16890  &  22393  &  36  &  510  &  511  \\ 
 53772-2351-616    &    4257  &  18800  &  23650  &  19  &  83  &  83  \\ 
 53794-1810-504    &    2991  &  14673  &  19637  &  -18  &  -344  &  -344  \\ 
 53794-2353-600    &    3115  &  10301  &  12216  &  92  &  2408  &  2901  \\ 
 53795-2216-298    &    3334  &  17934  &  22134  &  -16  &  -178  &  -178  \\ 
 53799-2222-388    &    3109  &  16042  &  21070  &  3  &  -567  &  -568  \\ 
 53800-2128-415    &    3260  &  11089  &  13105  &  106  &  2672  &  3172  \\ 
 53801-2428-356    &    4260  &  17357  &  21034  &  -49  &  -342  &  -342  \\ 
 53816-1843-008    &    3214  &  17328  &  22946  &  -9  &  -507  &  -507  \\ 
 53816-1843-502    &    3731  &  19401  &  23493  &  -36  &  -558  &  -559  \\ 
 53816-1843-584    &    3188  &  12348  &  15589  &  89  &  1960  &  1964  \\ 
 53818-2440-539    &    5832  &  18680  &  22599  &  40  &  368  &  368  \\ 
 53827-1713-063    &    3072  &  16202  &  19620  &  40  &  299  &  299  \\ 
 53827-1716-350    &    4097  &  19387  &  23656  &  -29  &  -257  &  -257  \\ 
 53828-2237-093    &    3131  &  16441  &  20726  &  0  &  220  &  220  \\ 
 53845-2233-196    &    3322  &  18757  &  23159  &  -44  &  -621  &  -620  \\ 
 53847-1775-420    &    2378  &  18475  &  25508  &  20  &  -45  &  -45  \\ 
 53851-1817-069    &    1992  &  12907  &  18676  &  40  &  1154  &  1154  \\ 
 53851-2023-065    &    2407  &  18911  &  25084  &  -12  &  -1364  &  -1364  \\ 
 53855-2491-252    &    3857  &  10980  &  13282  &  222  &  1729  &  1781  \\ 
 53858-2101-348    &    2364  &  15837  &  21926  &  11  &  -178  &  -178  \\ 
 53874-2152-108    &    4483  &  20349  &  25580  &  -5  &  -282  &  -282  \\ 
 53874-2173-354    &    4425  &  15002  &  18929  &  0  &  105  &  105  \\ 
 53875-2508-353    &    2609  &  12370  &  17747  &  -18  &  1368  &  1368  \\ 
 53877-2512-582    &    4530  &  14909  &  19081  &  53  &  1035  &  1037  \\ 
 53883-1793-573    &    2636  &  12781  &  16970  &  44  &  1612  &  1612  \\ 
 53884-1796-591    &    4871  &  15675  &  18800  &  161  &  746  &  746  \\ 
 53886-2164-503    &    3524  &  10830  &  12832  &  134  &  2278  &  2609  \\ 
 54084-2501-120    &    3469  &  12275  &  15254  &  116  &  1564  &  1565  \\ 
 54086-2420-240    &    2094  &  9655  &  11951  &  67  &  2842  &  3384  \\ 
 54095-2583-462    &    3192  &  16993  &  21672  &  1  &  -257  &  -256  \\ 
 54095-2583-615    &    2661  &  9378  &  11669  &  123  &  2019  &  2148  \\ 
 54097-2478-377    &    1889  &  8287  &  16515  &  23  &  262  &  262  \\ 
 54115-2102-018    &    3390  &  16757  &  21821  &  -21  &  -175  &  -175  \\ 
 54115-2493-278    &    3222  &  18640  &  33760  &  -19  &  -284  &  -5224  \\ 
 54138-1844-112    &    3773  &  12483  &  15143  &  193  &  1129  &  1129  \\ 
 54139-2425-442    &    4915  &  13077  &  15906  &  227  &  1465  &  1502  \\ 
 54144-1845-637    &    4159  &  15735  &  19109  &  -32  &  -279  &  -279  \\ 
 54149-2488-521    &    3240  &  8994  &  10569  &  154  &  1711  &  1958  \\ 
 54152-1803-323    &    3733  &  16984  &  21326  &  -15  &  -159  &  -158  \\ 
 54153-2242-199    &    3262  &  17364  &  21937  &  -59  &  -1563  &  -1563  \\ 
 54154-2606-614    &    3091  &  16227  &  20902  &  21  &  229  &  229  \\ 
 54175-1807-009    &    2546  &  14708  &  20815  &  2  &  1049  &  1049  \\ 
 54176-1847-630    &    3521  &  15330  &  19409  &  3  &  -261  &  -261  \\ 
 54176-2485-231    &    6537  &  21678  &  25958  &  -156  &  -602  &  -602  \\ 
 54178-2492-562    &    5245  &  19656  &  24010  &  -72  &  -524  &  -524  \\ 
 54178-2592-153    &    2310  &  17164  &  23418  &  7  &  -553  &  -552  \\ 
 54180-2509-244    &    3607  &  10674  &  12726  &  159  &  2145  &  2505  \\ 
 54207-2595-541    &    4036  &  15870  &  20642  &  41  &  830  &  831  \\ 
 54230-2641-570    &    3999  &  10882  &  13114  &  182  &  1742  &  1916  \\ 
 54232-2166-030    &    3003  &  17051  &  21841  &  67  &  3678  &  3678  \\ 
 54243-2518-374    &    3843  &  16080  &  19818  &  104  &  1230  &  1231  \\ 
 54266-1725-540    &    2636  &  17294  &  24090  &  9  &  1055  &  1055  \\ 
 54266-1791-046    &    2225  &  16680  &  24033  &  20  &  -1356  &  -1355  \\ 
 54272-2744-535    &    1978  &  14196  &  22208  &  -28  &  -686  &  -685  \\ 
 54448-2424-109    &    2688  &  15162  &  19486  &  14  &  -296  &  -297  \\ 
 54477-2645-133    &    2916  &  10247  &  13309  &  92  &  2049  &  2218  \\ 
 54479-2646-204    &    3146  &  11068  &  12876  &  127  &  2787  &  3139  \\ 
 54481-2613-620    &    2418  &  17430  &  23989  &  -23  &  1054  &  1054  \\ 
 54483-2615-257    &    2074  &  12173  &  19889  &  -30  &  -468  &  -468  \\ 
 54483-2615-404    &    2816  &  17082  &  22540  &  -25  &  230  &  229  \\ 
 54495-2647-160    &    4420  &  13133  &  15745  &  158  &  661  &  660  \\ 
 54497-2885-561    &    3315  &  20523  &  26586  &  16  &  334  &  334  \\ 
 54502-2617-287    &    3000  &  18933  &  26141  &  -14  &  1160  &  1160  \\ 
 54507-1795-509    &    3421  &  17250  &  21896  &  -15  &  -837  &  -837  \\ 
 54523-2151-527    &    2704  &  15579  &  21479  &  -33  &  -513  &  -514  \\ 
 54524-2783-432    &    4446  &  10601  &  12423  &  183  &  1538  &  1839  \\ 
 54527-2769-354    &    2665  &  11186  &  14880  &  69  &  2634  &  2777  \\ 
 54529-2772-038    &    2515  &  10850  &  16580  &  -8  &  1557  &  1563  \\ 
 54534-2774-051    &    6160  &  18785  &  22415  &  -116  &  -372  &  -372  \\ 
 54535-2764-053    &    2525  &  19802  &  25741  &  27  &  -481  &  -482  \\ 
 54562-2955-608    &    3898  &  16919  &  21764  &  -75  &  -1200  &  -1199  \\ 
 54563-2795-140    &    4374  &  18846  &  22901  &  -48  &  -279  &  -280  \\ 
 54567-2517-624    &    2748  &  16077  &  21982  &  2  &  311  &  312  \\ 
 54570-2522-131    &    4061  &  17375  &  22099  &  8  &  -514  &  -514  \\ 
 54582-2526-404    &    3370  &  17926  &  23874  &  -14  &  -328  &  -328  \\ 
 54584-2520-442    &    3932  &  12126  &  14496  &  554  &  1869  &  1870  \\ 
 54585-2529-300    &    2687  &  9822  &  12863  &  23  &  1699  &  1740  \\ 
 54589-2532-090    &    4495  &  17861  &  21944  &  69  &  695  &  695  \\ 
 54589-2970-373    &    3373  &  11772  &  14476  &  87  &  2681  &  3143  \\ 
 54590-2971-326    &    3282  &  14706  &  19455  &  32  &  805  &  806  \\ 
\end{longtable}

\pagebreak

\end{document}